\documentclass[letterpaper, 10 pt, conference]{ieeeconf}
\IEEEoverridecommandlockouts
\usepackage{amsmath} 
\usepackage{amssymb}  
\usepackage{algorithm}
\usepackage{algpseudocode}
\usepackage{xcolor}
\usepackage{graphicx}
\usepackage[caption=false,font=footnotesize]{subfig}

\usepackage{hyperref}
\usepackage{tikz}
\usepackage[caption=false,font=footnotesize]{subfig}
\usetikzlibrary{arrows.meta,positioning,calc,shapes.geometric}
\usetikzlibrary{calc,fit}

\newtheorem{lemma}{Lemma}
\newtheorem{assumption}{Assumption}
\newtheorem{definition}{Definition}

\newtheorem{theorem}{Theorem}
\newtheorem{remark}{Remark}

\usepackage{cite}
\usepackage{booktabs,tabularx}

\title{\LARGE \bf
Balancing Confidentiality and Attack Detectability in Networked Control Systems over Lossy Wireless Channels
}

\author{Mahyar Mohammadi$^{1}$, Subhrakanti Dey$^{2}$  and Arunava Naha$^{1}$    
\thanks{*This work was fully supported by the Wallenberg AI, Autonomous
Systems and Software Program (WASP) funded by the Knut and Alice Wallenberg Foundation}
\thanks{$^{1}$Mahyar Mohammadi and Arunava Naha are with Department of Electrical Engineering, Linköping University, 58183 Linköping, Sweden
        {\tt\small mahyar.mohammadi@liu.se},\quad{\tt\small arunava.naha@liu.se}}%
\thanks{$^{2}$Subhrakanti Dey is with Department of Electrical Engineering, Uppsala University, 75103 Uppsala, Sweden
        {\tt\small subhrakanti.dey@angstrom.uu.se}}%
}

\begin{document}

\maketitle
\thispagestyle{empty}
\pagestyle{empty}

\phantomsection\label{sec:abs}   
\begin{abstract}
We study the joint protection of a networked control system (NCS) against eavesdropping and data integrity attacks over a lossy wireless channel. The eavesdropper intercepts the transmitted sensor measurements and control inputs to reconstruct the system state. In contrast, the integrity attacker in this study replaces the sensor measurements with fabricated data to drive the system to undesired operating conditions. Although these two threats are typically studied separately, a wireless NCS should be protected against both simultaneously, and a defence against one affects the other. We employ an observation-encoding scheme that renders the eavesdropper's state-estimation error covariance unbounded under packet drops, and show that transmission energy governs how long this error remains bounded. In particular, lower transmission energy improves confidentiality. For integrity protection, we combine stochastic physical watermarking with a CUSUM-based sequential change detector. The proposed attack detection scheme accounts for intermittent communication and encoded measurements. The detection delay is governed by the Kullback--Leibler divergence (KLD) between the pre- and post-attack distributions of the test signal, and we show that a higher transmission energy increases the KLD and thereby reduces the detection delay. Furthermore, we derive analytical expressions for the eavesdropper's estimation error and the worst-case expected KLD, providing guidelines for selecting transmission energy to balance confidentiality and integrity protection and characterising the privacy--security trade-off. Finally, we run numerical simulations to validate our theoretical findings and demonstrate the privacy--security trade-off in practical scenarios.


\end{abstract}

\section{INTRODUCTION} \label{sec:intro}
Networked control systems (NCSs), in which sensing, computation and actuation are connected through communication networks, are common in several applications such as power systems, transportation and industrial automation \cite{sandberg2015cyberphysical,cardenas2008research,sandberg2022secure}. Wireless networks make such systems flexible and inexpensive to deploy, but they also introduce challenges such as channel unreliability, packet drops, interference, and energy limitations, which degrade estimation and control performance \cite{sinopoli2005lqg,liu2021remote}. Because most NCSs use off-the-shelf networking components, commodity software, and low-cost sensors with limited energy and computational capabilities, they are also prone to attacks. Attacks on NCSs can be broadly classified as passive or active. Passive attacks seek to extract information without modifying transmitted data or disrupting system operation. For example, an eavesdropper may intercept sensor measurements and control inputs and use knowledge of the system dynamics to estimate sensitive states \cite{dibaji2019systems,humayed2017cyber}. Unencrypted communication can facilitate such inference. Limited energy, computational resources and stringent latency requirements can, however, complicate the deployment of encryption \cite{darup2021encrypted}. On the other hand, active attacks manipulate data or disrupt system operation. These include data deception attacks, in which an adversary falsifies measurements or control commands to mislead the controller or alter the physical process. A replay attack is a form of data deception in which previously recorded, legitimate data are retransmitted as current data, potentially concealing malicious activity.  

Existing studies often address eavesdropping and data deception attacks separately. However, a defence against one threat can influence vulnerability to the other, creating potential trade-offs between privacy protection and attack resilience, as studied in  \cite{giraldo2017security,farokhi2018security}. These interactions motivate joint defences that address both threats while maintaining acceptable control performance, particularly in safety-critical applications. Developing such defences requires characterising their achievable performance bounds and understanding how design parameters influence the protection provided against each threat. Motivated by these challenges, we studied a joint defence mechanism against eavesdropping and data deception attacks in NCSs. Specifically, we consider measurement replacement attacks, a form of data deception in which an adversary spoofs sensor measurements by replacing legitimate data with falsified values.

\subsection{Related Work} \label{subsec:lit_reveiw}
\textbf{State Privacy.} 
Traditional cryptography-based methods can protect data confidentiality but can impose computational and energy costs on resource-constrained devices \cite{yan2024privacy,lee2010price}. This motivates approaches based on system dynamics and communication properties, including state encoding, transmission scheduling, and noise injection.

State-secrecy code-based approaches encode the current state or its local estimate using the information previously received by the legitimate estimator. When the eavesdropper misses a packet that is received by the legitimate estimator, it loses the reference information needed to decode subsequent transmissions \cite{tsiamis2019state}. Related schemes address acknowledgement manipulation \cite{lucke2022coding}, operation without acknowledgements \cite{kennedy2024innovation}, active eavesdroppers \cite{crimson2025remote}, and multi-sensor networks \cite{chen2023dynamic}. In a different approach, transmission scheduling controls when information is sent to balance legitimate estimation accuracy and confidentiality \cite{leong2018transmission,tsiamis2017state}. Furthermore, noise injection can obscure transmitted information \cite{wang2020optimal}, although it may also degrade legitimate estimation. With suitable channel knowledge and spatial dimensions, artificial noise can be placed in the null space of the legitimate channel\cite{leong2018use}, avoiding additional interference at that legitimate receiver while impairing eavesdropping \cite{yan2024privacy}. In our study, we build on this state-encoding idea and apply it to encode the remote sensors' observations sent to the state estimator.

\textbf{Data Deception Attacks.} Data deception attacks on cyber-physical systems, for example NCSs, where an  adversary manipulates sensor measurements and/or control signals, have been studied extensively in the literature \cite{pasqualetti2013attack,naha2023quickest}.
The ability of such an adversary to compromise a system within a limited number of sensor manipulations, where measurements can be replaced, deleted, or inserted, has also been characterised formally in \cite{duo2025multistep}. 
Since detectors that only observe the received measurements may fail to detect such attacks \cite{mo2009secure,naha2023quickest}, active schemes based on physical watermarking add a private random signal to the control input. Therefore, test signals, for instance innovation signal from a Kalman filter, not generated by the true plant then lack the expected correlation with the watermark, which changes the innovation statistics \cite{mo2009secure,mo2015physical,satchidanandan2016dynamic,liu2021active}. Alternatively, parity-space residuals have been shown to capture the effect of replayed input and output data and to distinguish it from system faults, with both passive and active designs to enhance detection \cite{zhao2024replay}.
Recent work has reduced the cost of watermarking, e.g., through event-based \cite{zhao2023stochastic} or parsimonious \cite{naha2023quickest} policies, and has combined it with moving-target defence \cite{liu2023proactive} or dynamic encryption \cite{li2023dynamic}. Moving-target defence has further been extended to a data-driven setting, where a time-domain estimate of the $\nu$-gap metric is used to separate stealthy attacks from faults in \cite{luo2025time}. Additionally, countermeasures that exploit asynchronous sampling have been proposed against stealthy attacks on aperiodic sampled-data systems in \cite{chen2025asynchronous}.  

To detect attacks with minimal delay, sequential change-detection tests such as the cumulative sum (CUSUM) procedure \cite{tartakovsky2014sequential} have been combined with watermarking \cite{naha2023parsimonious,naha2023quickest,naha2022sequential}. In particular, \cite{naha2023quickest} derived the conditional innovation distributions before and after a measurement replacement attack and used them in an optimal CUSUM test. These results, however, assume that all measurements are received, so that the Kalman filter operates in steady state and the innovation is stationary. Attack detectability and identifiability with asynchronous, delayed, and multi-rate measurements have been analysed using a lifted system model \cite{sun2025attack}, 
and watermarking under packet drops has been considered in \cite{weerakkody2017bernoulli}; however, neither has been combined with encoding or sequential detection.

Combining these methods with measurement encoding and sensor packet losses requires further analysis. Packet losses make the estimation error covariance depend on the reception history \cite{sinopoli2005lqg}. Encoding also links each decoded measurement to the last received packet, which may already be falsified. Therefore, an attack detector, which is combined with a defence mechanism against eavesdropping must account for both the reception history and the encoding dependencies.

\subsection{Our Contributions} \label{subsec:contributions}
In this paper, we study the joint protection of an NCS against eavesdropping and measurement replacement attacks over a lossy sensor--estimator link. To this end, we combine observation encoding and stochastic physical watermarking to enhance both confidentiality and attack detection. In the proposed scheme, the sensors encode their observations before transmission based on the last observation received by the legitimate user, and the legitimate user decodes them accordingly. Since the channel is lossy, the eavesdropper may miss a packet that the legitimate user receives, causing the eavesdropper to lose the reference needed for decoding. We refer to such events as critical events. Once such an event occurs, the eavesdropper's state-estimation error covariance grows without bound, while the legitimate estimator's error covariance remains bounded. Therefore, it is important to reduce the average time before a critical event occurs to enhance the overall confidentiality of the system. Our study shows how the transmission energy affects the time to the critical event, provided that the other factors influencing the channel quality remain constant. Note that, similar to several previous studies, our proposed defence mechanism does not require the system to be open-loop unstable to be effective. On the other hand, to detect measurement replacement attacks effectively, we design a CUSUM-based sequential detection scheme that uses the known watermark. To this end, we derive the conditional innovation distributions under nominal and attacked conditions, taking into account packet losses and encoding, and use them to update the CUSUM statistic only when a packet is received. To study the performance of the detector, we derive an analytical expression for the worst-case expected Kullback--Leibler divergence (KLD) between the nominal and attacked distributions, which serves as a measure of attack detectability. In other words, a lower KLD means a longer detection delay, which is undesirable. Furthermore, our study shows that the higher the transmission energy, the longer the average time to the critical event, but the higher the attack detectability, highlighting the trade-off between confidentiality and detection performance. Our main
contributions are the following:
\begin{enumerate}
    \item We combine observation encoding and stochastic physical watermarking to jointly address eavesdropping and measurement replacement attacks in networked control systems over lossy wireless channels.

\item We show that, after a critical event, the encoding makes the eavesdropper's state-estimation error covariance unbounded while keeping the legitimate estimator's error covariance bounded. We show how transmission energy affects the time to this critical event. 

    \item We derive the distributions of the innovation signal under nominal and attacked conditions under packet losses and encoding, and design a CUSUM-based sequential attack detection scheme.  We also derive an analytical expression for the worst-case expected KLD. These results guide transmission-energy selection to balance confidentiality and attack detection.

    \item Numerical simulations support the analysis and illustrate the confidentiality--detection trade-off.
\end{enumerate}

    

\subsection{Organisation of the Paper} \label{subsec:organization}
The paper is organised as follows. Section II presents the system and threat models. Section III introduces the encoding scheme and the watermarking-based CUSUM detector. Section IV analyses their performance and the resulting confidentiality–detection trade-off. Numerical results are reported in Section V, and Section VI concludes the paper.

\subsection{Notation} \label{subsec:notation}
We use the following notation throughout the paper. Vectors are column vectors and are denoted by bold lowercase letters, e.g., $\mathbf{x}$. Matrices are denoted by bold uppercase letters, e.g., $\mathbf{A}$. The transpose of a matrix $\mathbf{A}$ is denoted by $\mathbf{A}^\top$. The identity matrix of appropriate dimension is denoted by $\mathbf{I}$. The expectation operator is denoted by $\mathbb{E}[\cdot]$. The trace of a matrix $\mathbf{A}$ is denoted by $\mathrm{tr}(\mathbf{A})$. The spectral radius of a matrix $\mathbf{A}$ is denoted by $\rho(\mathbf{A})$. The notation $\mathbf{A}\succeq 0$ ($\mathbf{A}\succ 0$) indicates that $\mathbf{A}$ is positive semidefinite (positive definite). The notation $\mathcal{N}(\mathbf{m},\mathbf{P})$ denotes a Gaussian distribution with mean $\mathbf{m}$ and covariance $\mathbf{P}$. $|\cdot|$ denotes the determinant of a matrix.

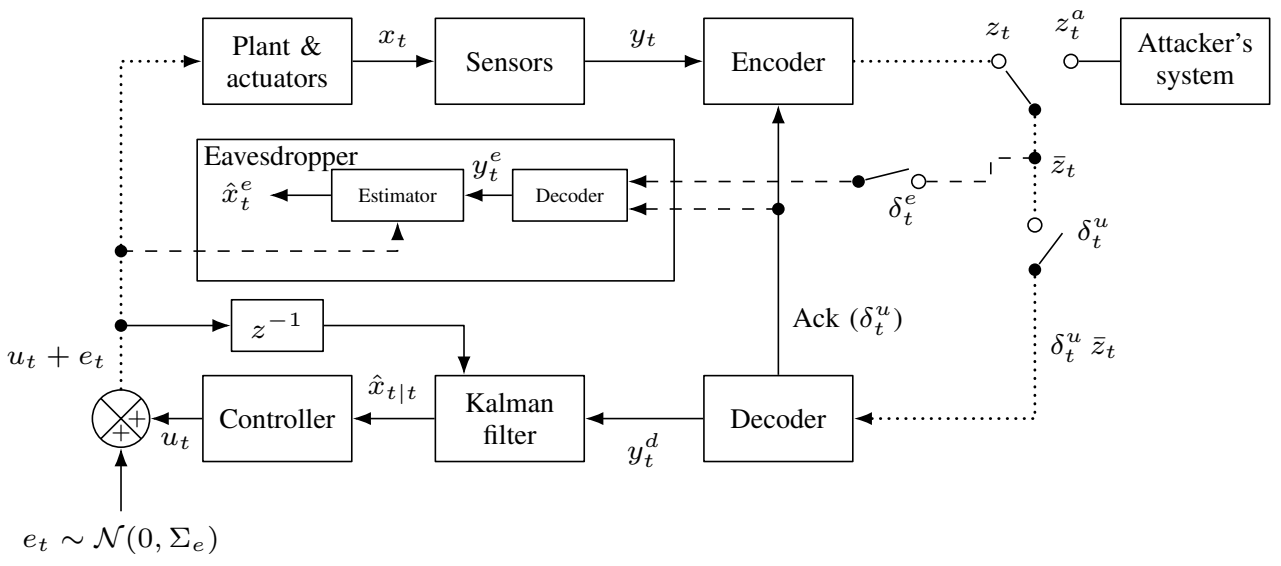
\begin{figure}[t]
  \centering
  \resizebox{\columnwidth}{!}{%
\begin{tikzpicture}[
    x=0.7cm, y=0.65cm,                 
    thin,
    line join=round,
    every node/.append style={font=\scriptsize},
    >={Latex[length=1.6mm,width=1.15mm]},
    block/.style   = {draw, rectangle, align=center, font=\scriptsize,
                      minimum width=1.3cm, minimum height=0.75cm, inner sep=1.5pt},
    sblock/.style  = {draw, rectangle, align=center, font=\tiny,
                      minimum height=0.45cm, inner sep=1.5pt},
    sum/.style     = {draw, circle, minimum size=5mm, inner sep=0pt,
                      path picture={\draw
                        (path picture bounding box.north west) --
                        (path picture bounding box.south east)
                        (path picture bounding box.north east) --
                        (path picture bounding box.south west);}},
    sign/.style    = {font=\tiny},
    contact/.style = {draw, circle, fill=white, minimum size=1.2mm, inner sep=0pt},
    joint/.style   = {circle, fill, minimum size=1.1mm, inner sep=0pt},
    note/.style    = {font=\scriptsize, inner sep=1.5pt},
    dots/.style    = {dash pattern=on 0pt off 2pt, line cap=round, line width=0.7pt},
  ]

  \node[block] (plant) at (1.95,4.8) {Plant \&\\ actuators};
  \node[block] (sens)  at (4.85,4.8) {Sensors};
  \node[block] (enc)   at (8.2,4.8)  {Encoder};
  \node[block] (dec)   at (8.2,0)    {Decoder};
  \node[block] (kf)    at (4.85,0)   {Kalman\\ filter};
  \node[block] (ctrl)  at (1.95,0)   {Controller};
  \node[block, minimum width=0.8cm, minimum height=0.45cm]
               (delay) at (1.95,1.25) {$z^{-1}$};          
  \node[block] (att)   at (13.4,4.8) {Attacker's\\ system};
  \node[sum]   (csum)  at (0,0) {};

  \node[contact] (zt)   at (10.95,4.8) {};
  \node[contact] (zta)  at (11.85,4.8) {};
  \node[joint]   (pole) at (11.4,4.15) {};
  \node[contact] (dc)   at (11.4,2.6)  {};
  \node[joint]   (piv)  at (11.4,2.0)  {};

  \draw[->] (plant.east) -- node[above] {$x_t$} (sens.west);

  \draw[->] (sens.east) -- node[above] {$y_t$} (enc.west);

  \draw[dots] (enc.east) -- (zt);
  \node[above] at (zt.north)  {$z_t$};
  \draw (att.west) -- (zta);
  \node[above] at (zta.north) {$z_t^{a}$};
  \draw (pole) -- ($(pole)!0.8!(zt)$);
  \draw[dots] (pole) -- node[right] {$\bar z_t$} (dc);
  \draw (piv) -- ++(55:0.6) node[right] {$\delta_t^u$};
  \draw[->, dots] (piv) |- node[pos=0.25, right] {$\delta_t^u\,\bar z_t$} (dec.east);

  \draw[->] (dec.west) -- node[below] {$y_t^d$} (kf.east);

  \draw[->] (dec.north) -- node[pos=0.2, right] {Ack ($\delta_t^u$)} (enc.south);

  \draw[->] (kf.west) -- node[above] {$\hat x_{t|t}$} (ctrl.east);
  \draw[->] (ctrl.west) -- node[below] {$u_t$} (csum.east);
  \node[sign] at ($(csum.center)+(0.2,0)$)  {$+$};
  \node[sign] at ($(csum.center)+(0,-0.2)$) {$+$};
  \draw[<-] (csum.south) -- ++(0,-0.85)
            node[below] {$e_t \sim \mathcal{N}(0,\Sigma_e)$};
  \coordinate (cbr) at (csum.north |- delay.west);
  \draw[->, dots] (csum.north) |- (plant.west);
  \node[left] at ($(csum.north)!0.5!(cbr)$) {$u_t + e_t$};
  \node[joint] at (cbr) {};
  \draw[->] (cbr) -- (delay.west);
  \draw[->] (delay.east) -| ([xshift=-4mm]kf.north);

  \draw (0.95,1.85) rectangle (6.9,3.75);
  \node[note, anchor=north west] at (0.95,3.75) {Eavesdropper};
  \node[sblock, minimum width=1.0cm]  (edec) at (5.6,3.0)  {Decoder};
  \node[sblock, minimum width=1.15cm] (eest) at (3.45,3.0) {Estimator};
  \draw[->] (edec.west) -- node[above] {$y_t^e$} (eest.east);
  \draw[->] (eest.west) -- (1.85,3.0) node[left] {$\hat x_t^{e}$};

  \coordinate (uetap) at (0,2.25);
  \node[joint] at (uetap) {};
  \draw[->, dashed] (uetap) -| (eest.south);

  \coordinate (ackin) at ([yshift=-1.2mm]edec.east);
  \coordinate (acktap) at (dec.north |- ackin);
  \node[joint] at (acktap) {};
  \draw[->, dashed] (acktap) -- (ackin);

  \coordinate (zin) at ([yshift=1.2mm]edec.east);
  \coordinate (ztap) at (11.4,3.5);
  \node[joint] at (ztap) {};
  \node[contact] (ezc) at (9.95,0 |- zin) {};
  \node[joint]   (ezp) at (9.2,0 |- zin)  {};
  \draw[dashed] (ztap) -| (10.85,0 |- zin) -- (ezc);
  \draw (ezp) -- ++(15:0.65);
  \node[below] at ($(ezp)!0.75!(ezc)$) {$\delta_t^e$};
  \draw[->, dashed] (ezp) -- (zin);
\end{tikzpicture}
}
\caption{The encoder sends
  $z_t = y_t - M^{t-t_k} y_{t_k}$ over a lossy channel ($\delta_t$), where an
  attacker may replace it with $z_t^a$. The decoder reconstructs $y_t^d$ for the
  Kalman filter, and the acknowledgement keeps both memories in sync.}
\label{fig:scheme1}
\end{figure}

\section{Preliminaries: System and Threat models} \label{sec:pre}
In this section, we formulate the problem of secure state estimation and attack detection in NCSs with lossy communication channels. We describe the system dynamics, the communication model, the encoding scheme, and the attack model.

In our study we considered a time-invariant linear stochastic system for modeling the plant dynamics, see Subsection~\ref{subsec:plant}. The sensor measurements are transmitted over a lossy communication channel to a remote estimator, which reconstructs the system state. The receiver (legitimate estimator) sends acknowledgements to the sensor. We further assume that the acknowledgements are reliably received by the sensor, since the controller--estimator block generally has much higher energy resources for communication. The estimator reconstructs the system state based on the received packets. The estimated states are then used for feedback control. The controller generates the control input based on the estimated state and transmits it to the actuator over a wireless network. An illustration of this setup is shown in Fig.~\ref{fig:scheme1} along with the eavesdropper and the attacker. We assume the controller has much higher energy resources compared to the sensor, allowing it to reliably transmit control signals. In other words, controller--actuator communication is assumed to be effectively lossless. Therefore, the primary focus of this work is on securing the sensor–estimator link against eavesdropping and attacks. A more general study considering attacks on the controller--actuator link, which is also lossy, is left for future work.

\subsection{Plant Model} \label{subsec:plant}
We consider a linear stochastic system for our study described by the following state-space equations
\begin{align}
  \mathbf{x}_{t+1} &= \mathbf{A} \mathbf{x}_t + \mathbf{B}\mathbf{u}_t + \mathbf{w}_t, \label{eq:plant}\\
  \mathbf{y}_t &= \mathbf{C} \mathbf{x}_t + \mathbf{v}_t, \label{eq:output}
\end{align}
where $\mathbf{x}_t\in\mathbb{R}^n$ is the state, $\mathbf{u}_t\in\mathbb{R}^m$ is the control input and
$\mathbf{y}_t\in\mathbb{R}^p$ is the measurement.

 
\begin{assumption}[Noises]\label{ass:noise}
$\mathbf{w}_t\sim\mathcal{N}(0,\mathbf{Q})$, $\mathbf{v}_t\sim\mathcal{N}(0,\mathbf{R})$ with $\mathbf{Q}\succeq 0$, and $\mathbf{R}\succ0$ are i.i.d.\ in time, mutually independent and
independent of $\mathbf{x}_0\sim\mathcal{N}(\hat{\mathbf{x}}_{0|0},\mathbf{P}_0)$.
\end{assumption}

\subsection{Lossy channel model} \label{subsec:lossy_channel}
As noted earlier, we assume only the sensor--estimator link is subject to packet losses.
Let $\delta_t\in\{0,1\}$ indicate successful packet reception at time $t$,
with $\delta_t=1$ if the packet is received and $\delta_t=0$ otherwise.
We model $\delta_t$ as a Bernoulli random variable with arrival probability
$\bar{\delta}$. Following~\cite{wu2015data}, we consider quadrature amplitude
modulation (QAM), and for a fixed transmit energy $E\geq 0$. Under this transmission scheme, $\bar{\delta}$ is
related to $E$ as
\begin{equation}\label{eq:energy}
  \bar{\delta} = 1-\epsilon^{E},\qquad
  \epsilon=\exp\Bigl(-\frac{\alpha}{N_0W_{\mathrm{bw}}}\Bigr)\in(0,1),
\end{equation}
where $N_0$ is the noise power spectral density, $W_{\mathrm{bw}}$ is the
channel bandwidth, and $\alpha\in(0,1)$ is a modulation-dependent constant.
Thus, achieving a target arrival probability $\bar{\delta}\in[0,1)$
requires the transmit energy
\[
  E(\bar{\delta})=\frac{\ln(1-\bar{\delta})}{\ln\epsilon}.
\]
We assume that the transmit energy is held fixed across transmissions and does not depend
on the data. We further assume that the packet-reception process is
independent of the transmitted content.

\subsection{Estimator and Controller} \label{subsec:estimator_controller}
The legitimate receiver employs a state estimator which predicts the state using the known system model and updates this prediction
whenever a measurement is received. Under nominal operation, the
predicted and filtered estimates are
\begin{align}
\hat{\mathbf{x}}_{t|t-1}
&= \mathbb{E}\bigl[\mathbf{x}_t\mid\mathcal{I}_{t-1}\bigr],
\label{eq:prediction_information}\\
\hat{\mathbf{x}}_{t|t}
&= \mathbb{E}\bigl[\mathbf{x}_t\mid\mathcal{I}_t\bigr].
\label{eq:filtering_information}
\end{align}
Here, $\mathcal{I}_t$ includes the initial information, received data, 
packet-reception history, and applied control inputs through time $t$. When $\delta_t=0$, no measurement
update is performed, so
$\hat{\mathbf{x}}_{t|t}=\hat{\mathbf{x}}_{t|t-1}$.

For the nominal LQG problem with packet losses,
the separation principle holds as studied in \cite{schenato2007foundations}.  Therefore, we use the
linear feedback law $\mathbf{u}_t=\mathbf{L}\hat{\mathbf{x}}_{t|t}$. The control input is transmitted over the lossless controller–actuator link.


\subsection{Eavesdropping model} \label{subsec:eavesdropping}
We consider an eavesdropper that can intercept packets transmitted over
both the sensor--estimator and controller--actuator links. The eavesdropper
knows the system model and uses its own estimator to reconstruct the
system state from the intercepted signals. The sensor--eavesdropper link is assumed to be lossy. Its packet-loss
process is independent of that of the legitimate sensor--estimator link,
and the two links may have different packet-arrival probabilities.

The defence mechanism aims to prevent accurate state reconstruction by
the eavesdropper despite its knowledge of the system model and access to
intercepted packets. We evaluate its effectiveness using the eavesdropper's
state-estimation error and the expected time until the eavesdropper loses
track of the system state.

\subsection{Attack model} \label{subsec:attack_model}
We consider a measurement replacement attack similar to that described in
\cite{naha2023quickest,liu2020secure}. The attacker aims to mislead the estimator
by replacing legitimate sensor packets with fabricated packets generated
by a stochastic dynamical model. The attack begins at an unknown time
$t_a$ and persists indefinitely.

For every $t\geq t_a$, the transmitted sensor packet is replaced by
$\bar{\mathbf{z}}_t=\mathbf{z}^a_t$, where
\begin{equation}\label{eq:attack}
  \mathbf{z}^a_{t+1}
  = \mathbf{A}_a\mathbf{z}^a_t+\mathbf{w}^a_t,
  \qquad
  \mathbf{w}^a_t\overset{\mathrm{i.i.d.}}{\sim}
  \mathcal{N}(0,\mathbf{Q}_a),
\end{equation}
with $\mathbf{Q}_a\succ0$ and $\rho(\mathbf{A}_a)<1$.
The attack process is initialised in its stationary distribution,
\[
  \mathbf{z}^a_{t_a}\sim\mathcal{N}(0,\boldsymbol{\Sigma}_z),
\]
where $\boldsymbol{\Sigma}_z\succ0$ is the unique solution to
\[
  \boldsymbol{\Sigma}_z
  =\mathbf{A}_a\boldsymbol{\Sigma}_z\mathbf{A}_a^\top+\mathbf{Q}_a.
\]
We assume the initial attack state is independent of the attack noise sequence
$\{\mathbf{w}^a_t\}_{t\geq t_a}$. The entire attack process is independent
of $\mathbf{x}_0$, $\mathbf{w}$, $\mathbf{v}$, $\mathbf{e}$, and $\delta$.
Fabricated packets are subject to the same packet-loss mechanism as
genuine packets.

\section{Proposed Defence Mechanisms} \label{sec:defence}
In this section, we propose defence mechanisms to protect the NCS against the measurement replacement attacks and ensure the privacy of the internal states of the system. First, we introduce an encoding-decoding scheme that helps maintain the confidentiality of the system states in the presence of eavesdroppers without degrading the performance of the legitimate estimator.

\subsection{Defence Against Eavesdropping: Observation Encoding-Decoding Scheme} \label{subsec:obs_encoding} 
The objective of the observation encoding-decoding scheme is to ensure that the legitimate estimator can accurately reconstruct the system states while preventing eavesdroppers from gaining useful information about the internal states of the system. 

Let $\tau_t:=\max\{s<t:\delta_s=1\}$ be the last reception time before $t$. The sensor encodes and transmits the following signal to the legitimate estimator:
\begin{equation}\label{eq:encoder}
  \mathbf{z}_t = \mathbf{y}_t - \mathbf{M}^{k_t} \mathbf{y}_{\tau_t}\text{, where } k_t = t - \tau_t.
\end{equation}
Here $\mathbf{M}\in\mathbb{R}^{p\times p}$ is a matrix with at least one eigenvalue strictly greater than 1 and is known to both the encoder and the legitimate estimator. 

Let $\delta_t^u\in\{0,1\}$ indicate whether the packet $\bar{\mathbf{z}}_t$ is received
by the legitimate estimator. The decoder then reconstructs
\begin{equation}\label{eq:decoder}
  \mathbf{y}^d_t = \delta_t^u\bigl(\bar{\mathbf{z}}_t + \mathbf{M}^{k_t} \mathbf{y}^d_{\tau_t}\bigr).
\end{equation}
For the legitimate estimator, this ensures that it can accurately reconstruct the system states as long as it receives the encoded packets. However, for an eavesdropper there could be a critical event say at time $t_c$ with it missing a packet but the legitimate estimator receiving it. Once this happens, the eavesdropper loses the reference $\mathbf{y}_{\tau_{t_c}}$ and can no longer reconstruct the subsequent measurements. Furthermore, note that without attack, $\bar{\mathbf{z}}_t=\mathbf{z}_t$. As shown in Lemma~\ref{lem:eaves}, the state estimation error of the eavesdropper grows unbounded after such a critical event, ensuring the confidentiality of the system states. On the other hand, the legitimate estimator continues to accurately reconstruct the system states as long as it receives the encoded packets.

\subsection{Defence Against Measurement Replacement Attacks} \label{subsec:meas_repl}
To protect the NCS against measurement replacement attacks, we propose a mechanism that allows the legitimate estimator to detect such attacks as early as possible. 

First, similar to several existing approaches, we add a physical watermarking signal $\bf{e}_t \sim \mathcal{N}(0, \mathbf{\Sigma}_e)$ to the control input. Specifically, the controller transmits 
\begin{equation}
  \mathbf{u}_t + \mathbf{e}_t.
  \label{eq:watermarking_signal}
\end{equation}
In our study, we take $\mathbf{e}_t$ to be a random signal known to the legitimate estimator but unknown to the attacker. As studied before \cite{naha2023parsimonious,naha2023quickest}, the addition of such a watermarking signal increases the detectability of data deception attacks.

\textbf{Test Signal.} We use a combination of the innovation signal (see \eqref{eq:innov}) from the state estimator and the watermarking signal as the test signal to detect measurement replacement attacks. The state estimation equations with the watermarking signal and lossy measurements are given as
\begin{align}
  &\hat{\mathbf{x}}_{t|t-1} = (\mathbf{A}+\mathbf{B}\mathbf{L})\hat{\mathbf{x}}_{t-1|t-1} + \mathbf{B}\mathbf{e}_{t-1}, \label{eq:kfpred}\\
  &\mathbf{P}_{t|t-1} = \mathbf{A}\mathbf{P}_{t-1|t-1}\mathbf{A}^\top + \mathbf{Q}, \label{eq:kfP}\\
  &\boldsymbol{\Gamma}_t = \mathbf{C}\mathbf{P}_{t|t-1}\mathbf{C}^\top + \mathbf{R},\qquad \mathbf{K}_t = \mathbf{P}_{t|t-1}\mathbf{C}^\top\boldsymbol{\Gamma}_t^{-1}, \label{eq:kfK}\\
  &\hat{\mathbf{x}}_{t|t} = \hat{\mathbf{x}}_{t|t-1} + \delta_t^u \mathbf{K}_t\boldsymbol{\gamma}_t, \label{eq:kfx}\\
  &\mathbf{P}_{t|t} = \mathbf{P}_{t|t-1}-\delta_t^u\mathbf{K}_t\mathbf{C}\mathbf{P}_{t|t-1}. \label{eq:kfupd}
\end{align}
Finally, the innovation signal is given as provided the packet arrives, 
\begin{equation}\label{eq:innov}
  \boldsymbol{\gamma}_t={\mathbf{y}}_t^d-\mathbf{C}\hat{\mathbf{x}}_{t|t-1}.
\end{equation}
Here $\mathbf{P}_{t|t-1} = \mathbb{E}[(\mathbf{x}_t-\hat{\mathbf{x}}_{t|t-1})(\mathbf{x}_t-\hat{\mathbf{x}}_{t|t-1})^\top \mid \mathcal{I}_{t-1}]$ is the predicted error covariance matrix, and $\mathbf{P}_{t|t} = \mathbb{E}[(\mathbf{x}_t-\hat{\mathbf{x}}_{t|t})(\mathbf{x}_t-\hat{\mathbf{x}}_{t|t})^\top \mid \mathcal{I}_t]$ is the updated error covariance matrix. \newline

\noindent\textbf{Test statistic.}
We use a CUSUM test statistic \cite{tartakovsky2014sequential} to detect the presence of a measurement replacement
attack. The statistic recursively accumulates the log-likelihood ratio
of the joint innovation and watermark signals under the attack
hypothesis relative to the nominal hypothesis. The recursion is given in
Theorem~\ref{thm:cusum_test}.

\begin{theorem}\label{thm:cusum_test}
Let Assumption~\ref{ass:noise} hold. The CUSUM test statistic to detect the measurement replacement attack~\eqref{eq:attack} under packet dropouts and the encoding--decoding scheme~\eqref{eq:encoder}--\eqref{eq:decoder} takes the following form:
\begin{align}
  S_t&=\max\{0,W_t\},\qquad W_0=0,\label{eq:cusum_S}\\
  W_t&=\begin{cases}
    \max\bigl\{W_{t-1}+\ell^{\mathrm{g}}_t,\;W_{t-1}+\ell^{\mathrm{f}}_t\bigr\}, & \delta^u_t=1,\\
    W_{t-1}, & \delta^u_t=0,
  \end{cases}\label{eq:cusum_W}
\end{align}
where, for $j\in\{\mathrm{g},\mathrm{f}\}$,
\begin{equation}\label{eq:cusum_llr}
  \ell^{j}_t=\ln\frac{f^{j}_{\tilde{\boldsymbol{\gamma}}_t}\bigl(\bar{\boldsymbol{\gamma}}_t\mid\mathcal{I}_{t-1}\bigr)}{f_{\boldsymbol{\gamma}_t}\bigl(\bar{\boldsymbol{\gamma}}_t\mid\mathcal{I}_{t-1}\bigr)}.
\end{equation}
Here $f^{\mathrm{f}}_{\tilde{\boldsymbol{\gamma}}_t}$ and $f^{\mathrm{g}}_{\tilde{\boldsymbol{\gamma}}_t}$ are the densities of the attacked innovation when the last received packet $\bar{\mathbf{z}}_{\tau_t}$ is falsified ($t_a\le\tau_t$) or genuine ($\tau_t<t_a$), respectively,
\begin{equation}\label{eq:cusum_H1}
  \bigl\{\tilde{\boldsymbol{\gamma}}_t\mid\mathcal{I}_{t-1},\delta_t^u=1\bigr\}
  \sim\mathcal{N}\bigl(\boldsymbol{\mu}^{j}_{t},\,\boldsymbol{\Sigma}^{j}_{t}\bigr),
\end{equation}
\begin{multline}\label{eq:cusum_mu}
  \boldsymbol{\mu}^{j}_{t}=
  \mathbf{M}^{k_t}\mathbf{y}^d_{\tau_t}-\mathbf{C}(\mathbf{A}+\mathbf{B}\mathbf{L})\hat{\mathbf{x}}_{t-1|t-1}-\mathbf{C}\mathbf{B}\mathbf{e}_{t-1}\\
  +\begin{cases}
    \mathbf{A}_a^{k_t}\bar{\mathbf{z}}_{\tau_t}, & j=\mathrm{f}\ (t_a\le\tau_t),\\
    \mathbf{0}, & j=\mathrm{g}\ (\tau_t<t_a),
  \end{cases}
\end{multline}
\begin{equation}\label{eq:cusum_Sig}
  \boldsymbol{\Sigma}^{j}_{t}=
  \begin{cases}
    \boldsymbol{\Lambda}_{k_t}:=\sum_{i=0}^{k_t-1}\mathbf{A}_a^{i}\mathbf{Q}_a(\mathbf{A}_a^{i})^\top, & j=\mathrm{f},\\
    \boldsymbol{\Sigma}_z, & j=\mathrm{g},
  \end{cases}
\end{equation}
and, in the absence of an attack,
\begin{equation}\label{eq:cusum_H0}
  \bigl\{\boldsymbol{\gamma}_t\mid\mathcal{I}_{t-1},\delta_t^u=1\bigr\}\sim\mathcal{N}(\mathbf{0},\boldsymbol{\Gamma}_t), \text{where } \boldsymbol{\Gamma}_t \text{ is given in ~\eqref{eq:kfK}}.
\end{equation}
\end{theorem}

\begin{proof}
The proof of Theorem~\ref{thm:cusum_test} is provided in Appendix~\ref{App:the1}.
\end{proof}

\subsection{Hypothesis testing framework} \label{sec:hypothesis_testing}
In this subsection, we formulate the detection of the measurement replacement attack as a hypothesis testing problem. At each time step $t$, we consider the following hypotheses:
\begin{itemize}
  \item $\mathcal{H}_0$: No attack is present, i.e., the received innovation $\boldsymbol{\gamma}_t$ follows the nominal distribution~\eqref{eq:cusum_H0}.
  \item $\mathcal{H}_1$: An attack is present, i.e., the received innovation $\tilde{\boldsymbol{\gamma}}_t$ follows the attacked distribution~\eqref{eq:cusum_H1}.
\end{itemize}   
We perform the following test based on the test statistic $S_t$ to decide between the two hypotheses:
\begin{equation}
  \text{Decide } \mathcal{H}_1 \text{ if } S_t \ge h, \quad \text{otherwise decide } \mathcal{H}_0,
\end{equation}
where $h>0$ is a predefined threshold. 
Under the CUSUM test, which requires a set of conditions to be satisfied (see \cite{tartakovsky2014sequential}), the optimal threshold is $\log(ARL_h)$, where $ARL_h$ denotes the user-defined threshold on the average run length to false alarm (ARL), see (\ref{eq:arl}).  In practice, $ARL_h$ is chosen based on the desired trade-off between detection speed and false alarm rate. 
\begin{equation}\label{eq:arl}
  ARL := \mathbb{E}\bigl[\inf\{t\ge 1: S_t \ge h\} \mid \mathcal{H}_0\bigr],
\end{equation}
which represents the expected time to false alarm under the null hypothesis.

However, the conditions for optimality of the CUSUM test may not be satisfied for the current detection problem, and the chosen threshold $h$ may only provide a suboptimal trade-off between detection speed and false alarm rate.

In practice, the attacker's model may be unknown. In such cases, a joint estimation and detection scheme can estimate the post-change distributions online while detecting attacks, as demonstrated in \cite{naha2023quickest}. The present study assumes a known attacker model to benchmark the performance of CUSUM-based detection, which is also the best-case scenario, and to investigate the fundamental trade-offs between confidentiality and attack detectability. Future work could extend the current framework to accommodate unknown or time-varying attacker models.

\section{Statistical Performance Analysis} \label{sec:stat_analysis}
In this section, we analyse the statistical performance of the proposed defence mechanisms, focusing on the trade-off between the defence against eavesdropping and attacks. We derive analytical expressions for key performance metrics, such as the average time before the critical event, estimation error covariance for the eavesdropper and the legitimate receiver, the worst-case expected KLD, and other relevant statistical measures.

\subsection{Performance Metrics for defence against eavesdropping}
Our objective is to achieve perfect secrecy as defined in Definition~\ref{def:secrecy}. We adopt the notion of perfect secrecy of~\cite{tsiamis2019state}, adapted to the transmission of encoded measurements.
\begin{definition}[Perfect secrecy]\label{def:secrecy}
An encoding--decoding scheme~\eqref{eq:encoder}--\eqref{eq:decoder} achieves perfect
secrecy if and only if both of the following conditions hold:
\begin{enumerate}
  \item[(i)] the legitimate user's performance is optimal, i.e., for every $t$,
  \begin{equation*}
    \hat{\mathbf{x}}^u_{t|t}=\mathbb{E}\bigl[\mathbf{x}_t\mid\mathcal{I}_t\bigr],\qquad
    \mathbf{P}^u_{t|t}=\mathrm{Cov}\bigl(\mathbf{x}_t\mid\mathcal{I}_t\bigr),
  \end{equation*}
  where $\mathcal{I}_t$ is the information set in~\eqref{eq:filtering_information}. In other words, the encoding does not degrade the user's estimate.
  \item[(ii)] the eavesdropper's error diverges to infinity with probability one,
  \begin{equation*}
    \operatorname{tr}\bigl(\mathbf{P}^e_{t|t}\bigr)\xrightarrow{\ \mathrm{a.s.}\ }\infty
    \quad\text{as } t\to\infty .
  \end{equation*}
\end{enumerate}
\end{definition}
Next,  we introduce the key assumptions and metrics used to evaluate the performance of the defence mechanisms against eavesdropping. 
\begin{assumption}[Reception processes]\label{ass:reception}
Each packet is transmitted with a constant energy $E>0$. The reception
indicators $\delta^u_t,\delta^e_t\in\{0,1\}$ at the user and at the
eavesdropper satisfy:
\begin{enumerate}
  \item the pairs $\{(\delta^u_t,\delta^e_t)\}_{t\ge 1}$ are i.i.d.\ over $t$;
  \item for each $t$, $\delta^u_t$ and $\delta^e_t$ are independent, with
  \begin{equation*}
    \mathbb{P}\{\delta^u_t=1\}=\bar\delta_u, \quad
    \mathbb{P}\{\delta^e_t=1\}=\bar\delta_e.
  \end{equation*}
  \item we assume $\delta_u$ and $\delta_e$ always $> \bar{\delta}_c$, where $\bar{\delta}_c = \max(0.5, \tilde{\delta}_c)$, where $\tilde{\delta}_c$ is the critical arrival probability required for system's stability \cite{schenato2007foundations}.
\end{enumerate}
\end{assumption}

\begin{theorem}[Average time before the critical event]\label{thm:atbce}
Let Assumption~\ref{ass:reception} hold with $\bar\delta_u\in(0,1)$ and
$\bar\delta_e\in(0,1)$. Define the critical event and its first occurrence time
\begin{equation*}
  \mathcal{C}_t:=\{\delta^u_t=1,\ \delta^e_t=0\},\qquad
  t_c:=\min\{t\ge 1:\ \mathcal{C}_t \text{ occurs}\},
\end{equation*}
and let $p_c:=\bar\delta_u(1-\bar\delta_e)\in(0,1)$. 
The average time before the critical event (ATBCE) is
\begin{equation}\label{eq:atbce}
  \mathbb{E}[t_c]=\frac{1}{\bar\delta_u(1-\bar\delta_e)}.
\end{equation}
Assuming, $\bar \delta_u = \bar \delta_e$, the ATBCE relates to the transmission energy $E$ as
\begin{equation*}
  \mathbb{E}[t_c]=\frac{1}{\epsilon^E(1-\epsilon^E)},
\end{equation*}
where $\epsilon$ is given in \eqref{eq:energy}.
\end{theorem}

\begin{proof}
The proof of Theorem~\ref{thm:atbce} is provided in Appendix~\ref{App:the2}.
\end{proof}

\begin{assumption}[Eavesdropper]\label{ass:eaves}
The eavesdropper knows $(\mathbf{A},\mathbf{B},\mathbf{C},\mathbf{Q},\mathbf{R},\mathbf{M})$
and overhears the acknowledgements and the applied input $\mathbf{u}_t+\mathbf{e}_t$;
hence it knows $\tau_t$ and $t_c$. The encoder, the user and the eavesdropper share
the initial reference $\mathbf{y}_0$, i.e., $\tau_t:=0$ if $\delta^u_s=0$ for all
$1\le s<t$, and $\mathbf{r}^e_0:=\mathbf{y}_0$. At every reception ($\delta^e_t=1$)
the eavesdropper applies the legitimate decoder to its own references,
\begin{equation}\label{eq:eaves_dec}
  \mathbf{y}^e_t=\mathbf{z}_t+\mathbf{M}^{k_t}\mathbf{r}^e_{\tau_t},
\end{equation}
and feeds $\mathbf{y}^e_t$ to the Kalman filter~\eqref{eq:kfpred}--\eqref{eq:kfupd}
with $\delta^u_t$ replaced by $\delta^e_t$, producing $\hat{\mathbf{x}}^e_{t|t-1}$,
$\hat{\mathbf{x}}^e_{t|t}$. At every $t$ with $\delta^u_t=1$ it stores the reference
\begin{equation*}
  \mathbf{r}^e_t=\begin{cases}
    \mathbf{y}^e_t, & \delta^e_t=1,\\
    \mathbf{C}\hat{\mathbf{x}}^e_{t|t-1}, & \delta^e_t=0.
  \end{cases}
\end{equation*}
\end{assumption}
\begin{lemma} \label{lem:eaves}
Let $\rho(\mathbf{M})>1$ and let Assumptions~\ref{ass:noise}-~\ref{ass:eaves} hold. Define the
state estimation error covariances of the legitimate user and the eavesdropper as
$\mathbf{P}^u$ and $\mathbf{P}^e$.
Then the following statements hold.

\begin{enumerate}
  \item[(i)] \emph{Legitimate user.} The legitimate user's state estimation will satisfy the perfect secrecy condition as stated in Definition~\ref{def:secrecy}.
  \item[(ii)] \emph{Eavesdropper.} Suppose the critical event, as
  defined in Theorem~\ref{thm:atbce}, occurs at $t_c$ and that $\delta^e_t=1$ for
  all $t>t_c$, which is the best-case scenario for the eavesdropper. Let $(\mathbf{A},\mathbf{C})$ be detectable,
  $(\mathbf{A},\mathbf{Q}^{1/2})$ stabilisable, and
  $\mathbf{C}\mathbf{Q}\mathbf{C}^\top\succ0$, and let
  $\mathbf{P}^e_{t|t}:=\mathbb{E}\bigl[\tilde{\mathbf{x}}^e_{t|t}\tilde{\mathbf{x}}^{e\top}_{t|t}\bigr]$,
  with $\tilde{\mathbf{x}}^e_{t|t}:=\mathbf{x}_t-\hat{\mathbf{x}}^e_{t|t}$, be the
  actual error covariance of the eavesdropper's estimate. Then there exist a constant
  $c>0$, independent of $t_c$, and a finite $T_0$ such that
  \begin{equation*}
    \operatorname{tr}\bigl(\mathbf{P}^e_{t|t}\bigr)\ \ge\ c\,\rho(\mathbf{M})^{2(t-t_c)},
    \qquad t\ge t_c+T_0 .
  \end{equation*}
  Hence $\operatorname{tr}\bigl(\mathbf{P}^e_{t|t}\bigr)\to\infty$ as $t\to\infty$.
\end{enumerate}
\end{lemma}
\begin{proof}
Part~(i) follows from~\cite{schenato2007foundations}, since the user
decodes every received packet exactly. Part~(ii) is proved in
Appendix~\ref{App:lem_eaves}.
\end{proof}

\subsection{Performance Metrics for defence against attacks} 
Here our objective is to detect the attack as soon as possible after it occurs. In other words, we would like to have a KLD between the post-attack and pre-attack distributions that is as large as possible. Under the classical i.i.d. assumptions, the worst-case average detection
delay (WADD) of the optimal CUSUM test scales asymptotically as the detection
threshold divided by the KLD between the
post- and pre-attack distributions \cite{tartakovsky2014sequential}.
Applying this relationship to our setting, however, requires regularity
conditions that have not been established. Nevertheless, KLD remains a
useful measure of the statistical distinguishability of nominal and attacked
signals. In our setting, packet losses make the state-estimation error
covariance depend on the reception history, rather than converge to a
deterministic steady-state value. Consequently, the conditional KLD also
depends on this history, making its expectation difficult to evaluate
analytically. We therefore use the worst-case expected KLD as a tractable,
conservative measure of attack detectability, without asserting a
corresponding bound on WADD.
\begin{remark}\label{disc:wkld}
Under the persistent measurement replacement attack, the first packet received
at or after $t_a$ is decoded using a genuine reference
($\tau_t<t_a$, $j=\mathrm{g}$). All subsequent receptions use a reference
obtained from a falsified packet ($\tau_t\geq t_a$, $j=\mathrm{f}$).
The following KLD analysis therefore focuses on $j=\mathrm{f}$, which applies
after the first attacked reception. The detector still evaluates both
likelihoods because the attack onset is unknown.
\end{remark}
\begin{theorem}[Worst-case expected KLD]\label{thm:wkld}
Let Assumptions~\ref{ass:noise}-~\ref{ass:eaves} hold and $(\mathbf{A},\mathbf{C})$ be detectable and
$(\mathbf{A},\mathbf{Q}^{1/2})$ stabilisable.
Suppose that the last received packet is falsified, i.e., $t_a \le \tau_t$
($j=f$). Then the expected conditional KLD between the post- and pre-attack
conditional distributions used in the CUSUM test in Theorem~\ref{thm:cusum_test}, with
$f^f_{\tilde{\gamma}_t}$ and $f_{\gamma_t}$ given in \eqref{eq:cusum_H1} and \eqref{eq:cusum_H0}, and with the convention that the conditional KLD is zero when
$\delta^u_t=0$ (no packet is received and the CUSUM statistic~\eqref{eq:cusum_W}
is not updated), satisfies in steady state
\begin{equation}
\mathbb{E}\big[ D\big( f^f_{\tilde{\gamma}_t}, f_{\gamma_t} \mid I_{t-1} \big) \big]
 \ge \mathrm{WKLD}_f, \label{eq:wkld_bound}
\end{equation}
where the worst-case expected KLD, $\mathrm{WKLD}_f$, is given by
\begin{multline}
\mathrm{WKLD}_f = \frac{\bar{\delta}^u}{2} \Big(
 \operatorname{tr}\big(
 (\mathbf{C}\bar{\mathbf{P}}\mathbf{C}^\top + \mathbf{R})^{-1}
 (\mathbf{Q}_a + \mathbf{N}) \big) \\
 - p + \log \frac{\big| \mathbf{C}\mathbf{P}^{\star}\mathbf{C}^\top + \mathbf{R} \big|}
 {|\boldsymbol{\Sigma}_z|} \Big), \label{eq:wkld}
\end{multline}
where $\mathbf{N} := \mathbf{C}\mathbf{B}\boldsymbol{\Sigma}_e\mathbf{B}^\top\mathbf{C}^\top \succeq 0$. 
Here, $\bar{\mathbf{P}}$ is the stabilising solution of the MARE
\begin{equation*}
\bar{\mathbf{P}} = \mathbf{A}\bar{\mathbf{P}}\mathbf{A}^\top + \mathbf{Q} - \bar{\delta}^u \mathbf{A}\bar{\mathbf{P}}\mathbf{C}^\top (\mathbf{C}\bar{\mathbf{P}}\mathbf{C}^\top + \mathbf{R})^{-1} \mathbf{C}\bar{\mathbf{P}}\mathbf{A}^\top,
\end{equation*}
which satisfies $\limsup_{t\to\infty}\mathbb{E}[\mathbf{P}_{t|t-1}]\preceq\bar{\mathbf{P}}$~\cite{schenato2007foundations},
and $\mathbf{P}^{\star}$ is the stabilising solution of the standard ARE
(the MARE with $\bar{\delta}^u=1$)
\begin{equation*}
\mathbf{P}^{\star} = \mathbf{A}\mathbf{P}^{\star}\mathbf{A}^\top + \mathbf{Q} - \mathbf{A}\mathbf{P}^{\star}\mathbf{C}^\top (\mathbf{C}\mathbf{P}^{\star}\mathbf{C}^\top + \mathbf{R})^{-1} \mathbf{C}\mathbf{P}^{\star}\mathbf{A}^\top,
\end{equation*}
which is the error covariance of the lossless Kalman filter and lower bounds
$\mathbf{P}_{t|t-1}$ for every packet-reception realization, i.e.,
$\mathbf{P}^{\star}\preceq\mathbf{P}_{t|t-1}$ almost surely in steady state.
\end{theorem}

\begin{proof}
The proof of Theorem~\ref{thm:wkld} is provided in Appendix~\ref{App:the3}.
\end{proof}

\begin{remark}
The conditional KLD
$D\bigl(f^{j}_{\tilde{\boldsymbol{\gamma}}_t},f_{\boldsymbol{\gamma}_t}\mid\mathcal{I}_{t-1}\bigr)$
is a random variable, since $\boldsymbol{\Gamma}_t$, $\boldsymbol{\mu}^j_t$ and
$\boldsymbol{\Sigma}^j_t$ depend on the random packet-reception scenario. As a result,
its expected value has no closed-form expression. The WKLD in
Theorem~\ref{thm:wkld} overcomes this by providing a closed-form bound that
depends only on the system matrices and on the parameters $\bar{\delta}$
and $\boldsymbol{\Sigma}_e$, and can therefore be used to select the transmission
energy and the watermark.
\end{remark}
\begin{remark}[Dependence on transmission energy]\label{rem:wkld_energy}
By~\eqref{eq:energy}, $\bar{\delta}^u=1-\epsilon^{E}$ increases with $E$, and
$\bar{\mathbf{P}}$ is nonincreasing in
$\bar{\delta}^u$~\cite{schenato2007foundations}. Hence the trace term of~\eqref{eq:wkld},
$\operatorname{tr}\bigl((\mathbf{C}\bar{\mathbf{P}}\mathbf{C}^\top+\mathbf{R})^{-1}(\mathbf{Q}_a+\mathbf{N})\bigr)$,
which carries the watermark contribution $\mathbf{N}$, is nondecreasing in $E$. A
higher transmission energy thus makes the watermark more visible in the innovation. The log-determinant term of~\eqref{eq:wkld} depends only on
$\mathbf{P}^{\star}$ and $\boldsymbol{\Sigma}_z$ and is therefore independent of $E$.
Consequently, the bracketed term in~\eqref{eq:wkld} is nondecreasing in $E$, and
whenever it is positive, i.e., whenever the bound is informative, $\mathrm{WKLD}_f$
itself is nondecreasing in $E$ through both the prefactor $\bar{\delta}^u$ and the
trace term. The resulting trade-off with the ATBCE of Theorem~\ref{thm:atbce} is
evaluated numerically in Section~\ref{sec:numeric}.
\end{remark}

\section{NUMERICAL RESULTS} \label{sec:numeric}

\begin{figure}[t]
    \centering
    \subfloat[\label{fig:trace}]{%
        \includegraphics[width=0.5\linewidth]{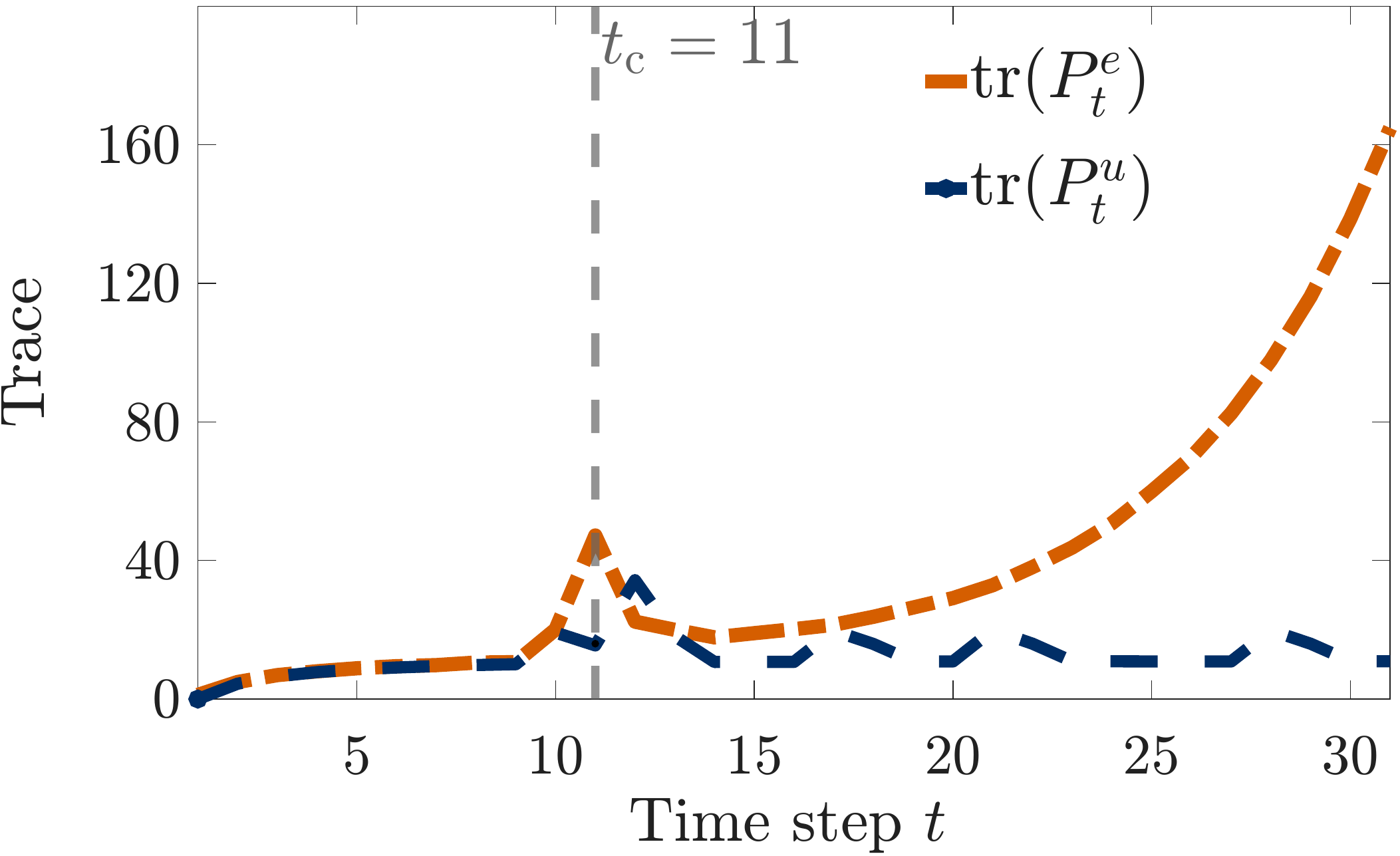}}
    \hfill
    \subfloat[\label{fig:second2}]{%
        \includegraphics[width=0.49\linewidth]{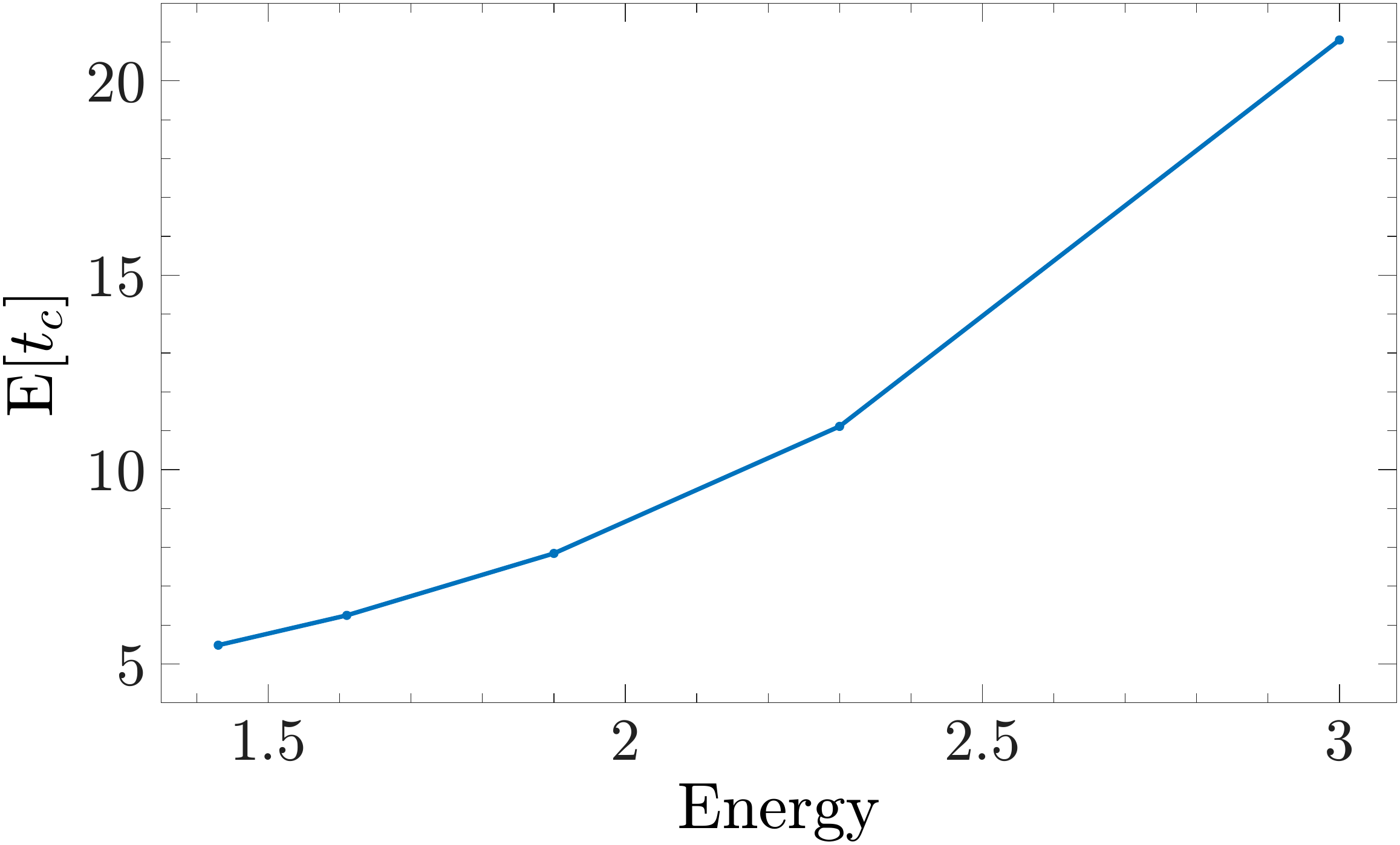}}
    
    \caption{(a) Trace of covariances
    (b)  $E[t_c]$ vs. Energy.}
    \label{fig:trace_both}
\end{figure}

\begin{figure}[t]
    \centering
    \subfloat[\label{fig:waddE}]{%
        \includegraphics[width=0.52\linewidth]{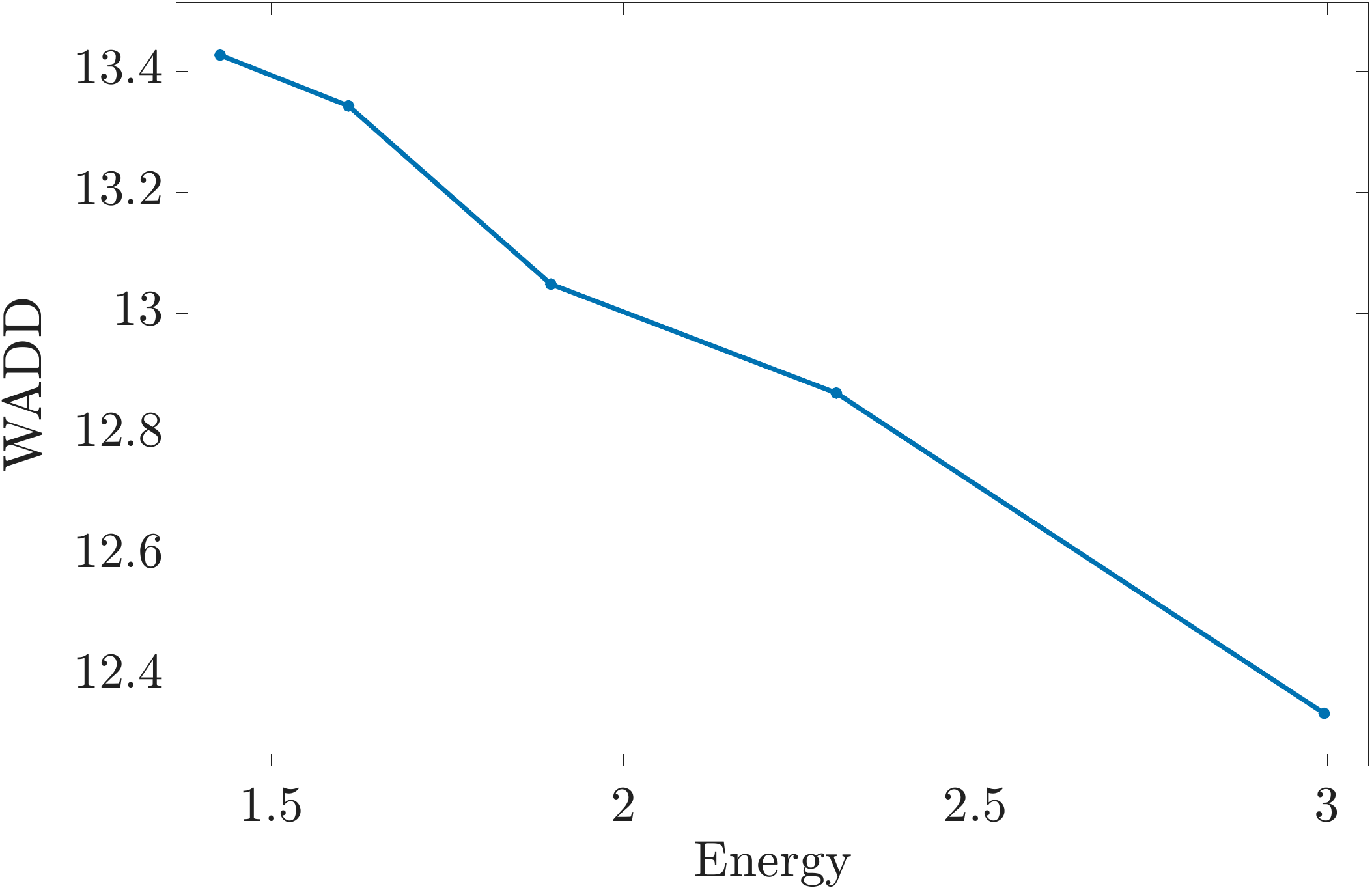}}
    \hfill
    \subfloat[\label{fig:wkldE}]{%
        \includegraphics[width=0.47\linewidth]{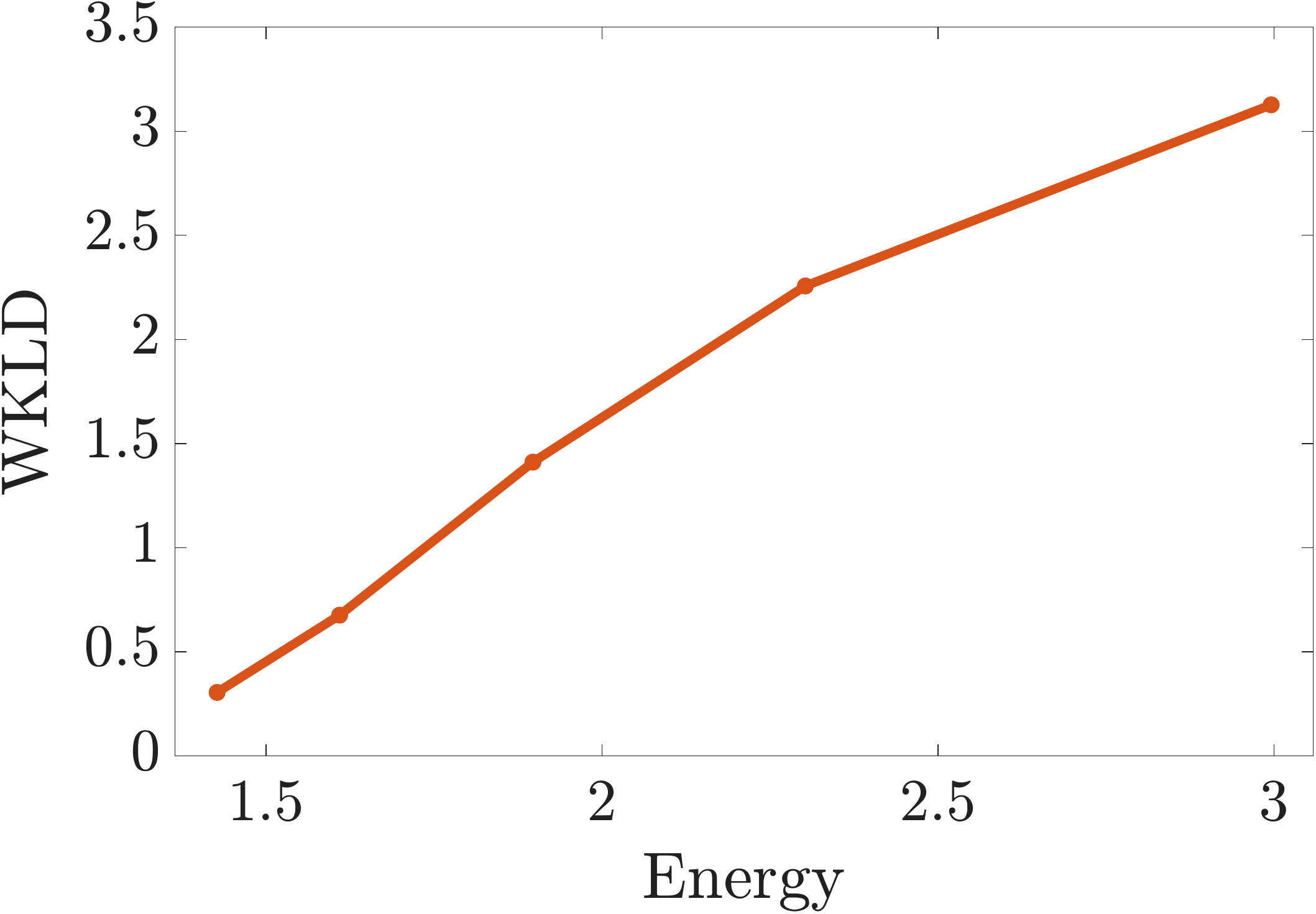}}
    \caption{(a) WADD vs.  Energy
    (b) WKLD vs. Energy.}
    \label{fig:trace_both}
\end{figure}

\begin{figure}[t]
    \centering
    \subfloat[\label{fig:trace2}]{%
        \includegraphics[width=0.49\linewidth]{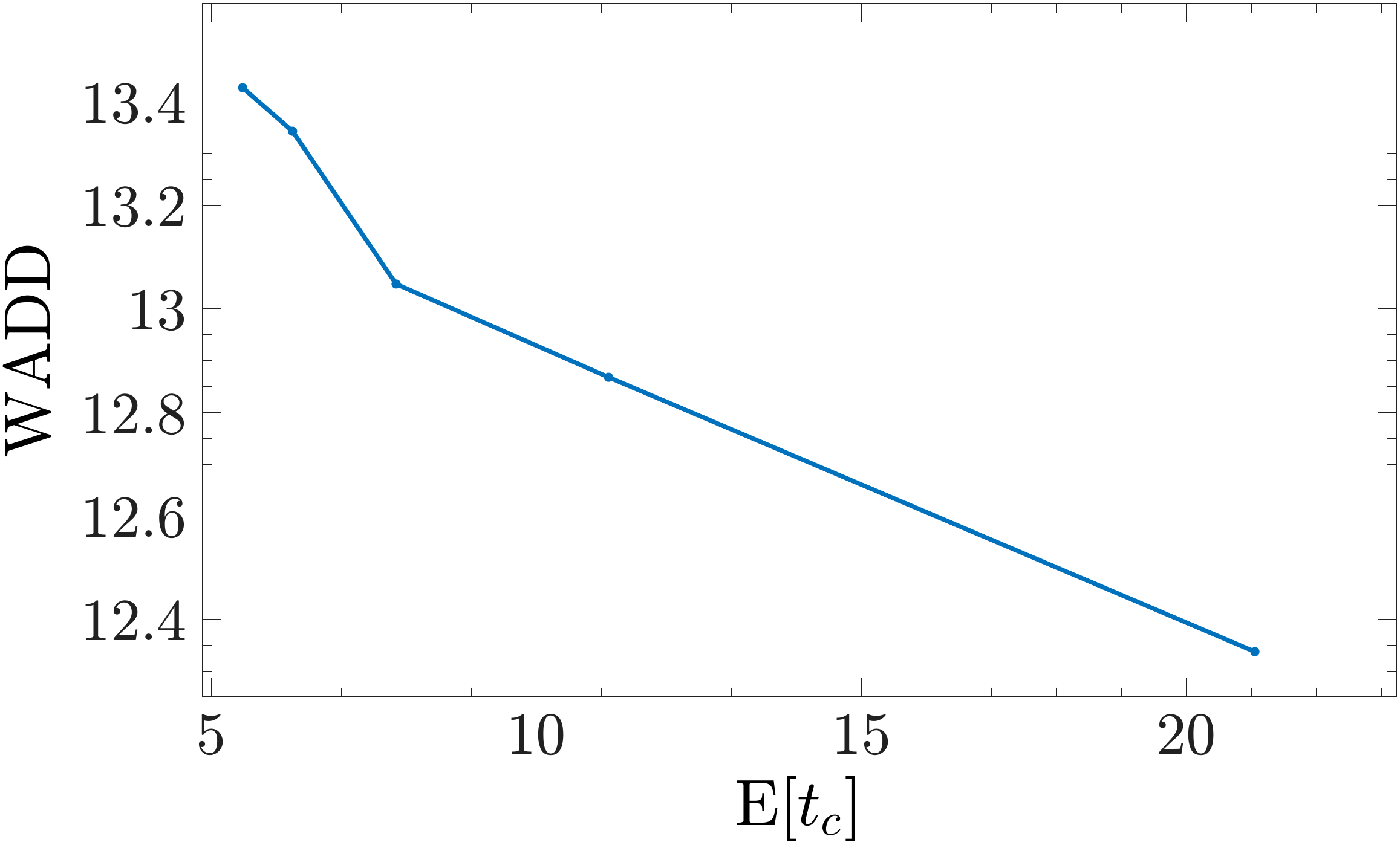}}
    \hfill
    \subfloat[\label{fig:second}]{%
        \includegraphics[width=0.47
        \linewidth]{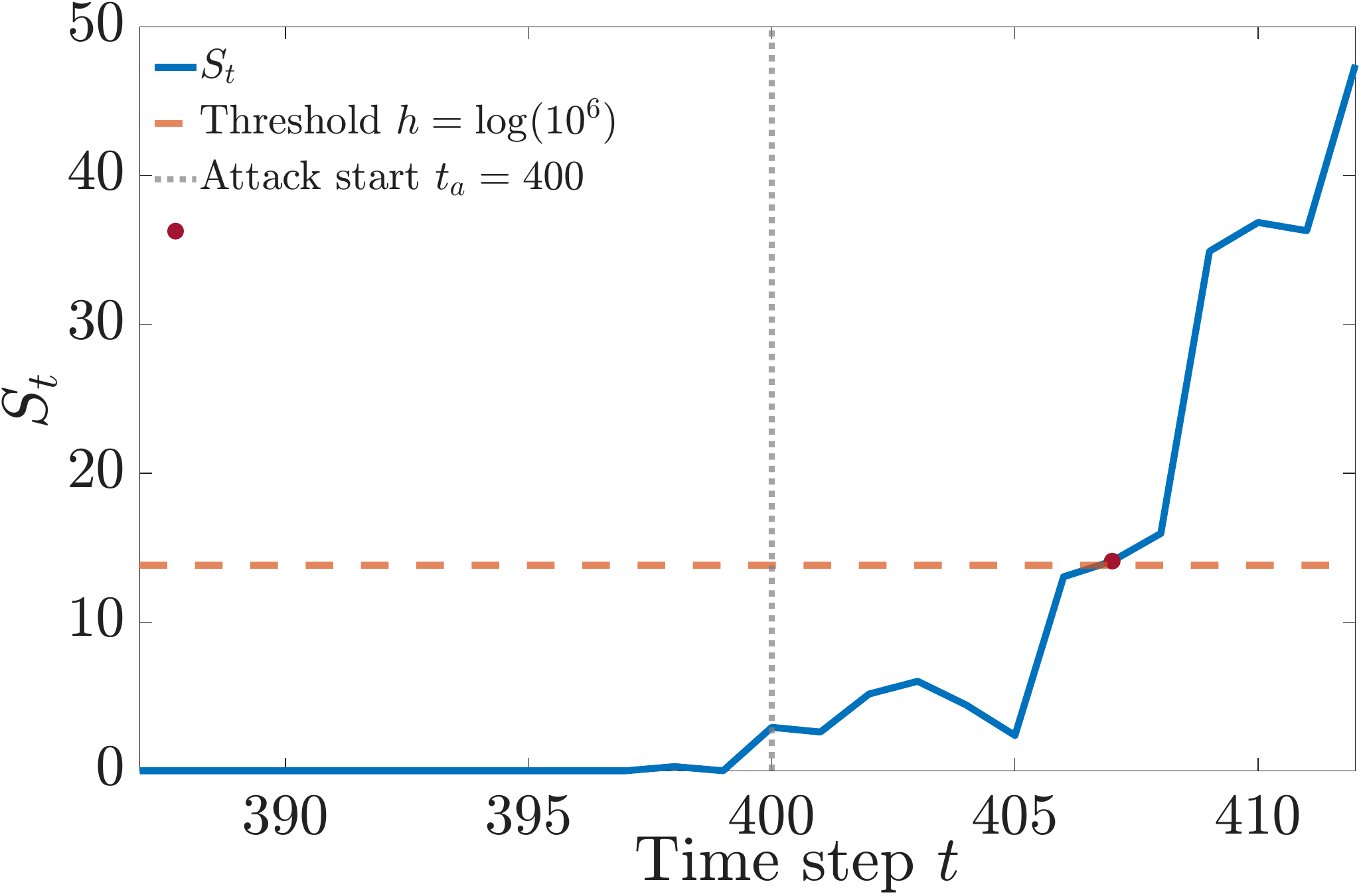}}
    \caption{(a) WADD vs.  $E[t_c]$
    (b) CUSUM Test.} 
    \label{fig:trace_both}
\end{figure}

\begin{figure}[h!]
  \centering
  \includegraphics[width=\columnwidth]{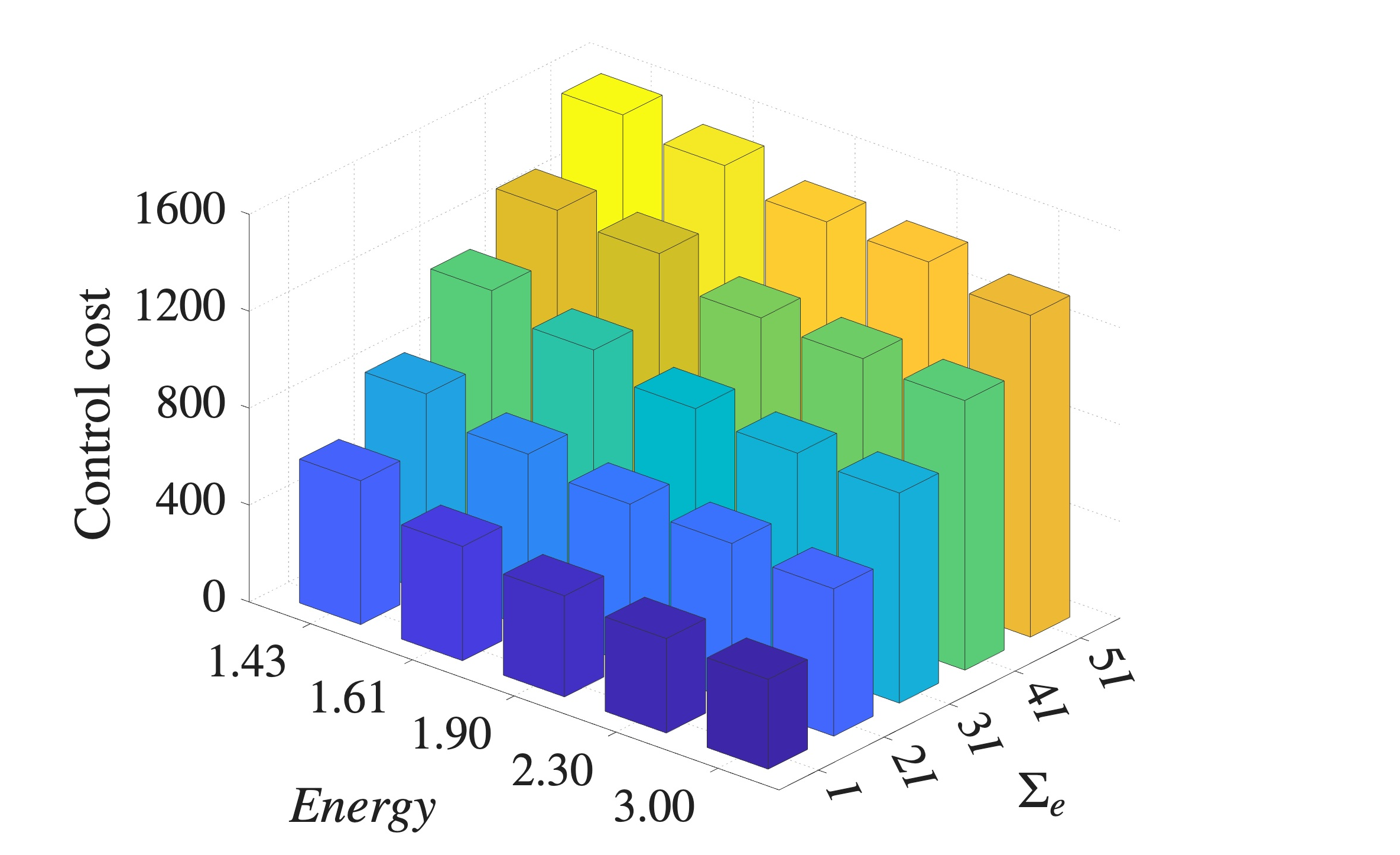}
  \caption{Control cost($J$) vs. noise covariance($\Sigma_e$) vs. Energy.}
  \label{fig:energy}
\end{figure}

In this section, we evaluate the proposed defence mechanisms through
Monte Carlo simulations. We first examine the effect of observation
encoding on the state estimates of the legitimate user and the
eavesdropper. We then assess the watermarking-based CUSUM detector, study
how the transmission energy affects attack detectability and
confidentiality, and finally quantify the control cost of the watermark and its relationship with transmission energy.

\subsection{Simulation Setup}\label{subsec:setup}
 
We consider a sixth-order linear stochastic system~\eqref{eq:plant}--\eqref{eq:output}, with the parameters
listed in Appendix~\ref{apdx:num_parameters}. 
The legitimate receiver uses the estimator and controller described in
Sections~\ref{sec:pre} and~\ref{sec:defence}. Observations are encoded according to~\eqref{eq:encoder}, with
$\rho(\mathbf{M})>1$, and the controller adds an independent Gaussian
watermark $\mathbf{e}_t\sim\mathcal{N}(\mathbf{0},\boldsymbol{\Sigma}_e)$
to the applied input.
 
Packet reception follows~\eqref{eq:energy} with $\alpha/(N_0W_{bw})=1$, $i.e.$,
$\epsilon=e^{-1}$, so that $\bar{\delta}=1-e^{-E}$. We consider the
transmission energies
$E\in\{1.43,\,1.61,\,1.90,\,2.30,\,3.00\}$, which correspond to the
arrival probabilities
$\bar{\delta}\in\{0.76,\,0.80,\,0.85,\,0.90,\,0.95\}$. The legitimate and
eavesdropper reception processes are independent and have equal arrival
probabilities, $\bar{\delta}^u=\bar{\delta}^e=\bar{\delta}$, unless
otherwise stated. 
 
The replacement attack starts at $t_a$ and follows~\eqref{eq:attack}, with
$\rho(\mathbf{A}_a)<1$ and stationary initialisation. The fabricated packets
experience the same packet losses as genuine packets. The CUSUM
threshold is $h=\log(\mathrm{ARL}_h)$ with $\mathrm{ARL}_h=10^6$, i.e.,
$h\approx 13.8$. Results are averaged over 1000
independent trials. 
Comparisons between schemes use matched noise and reception realisations.

\subsection{Confidentiality under Observation Encoding}\label{subsec:conf}
 
Fig.~\ref{fig:trace} shows the trace of ${\bf{P}}_{t|t}^u$ from the Kalman filter of a single trial and estimated ${\bf{P}}_{t|t}^e$ over 1000 Monte Carlo trials, where packet received time instances are kept fixed.
 
Before the critical event, every packet received by the user is also
received by the eavesdropper. The eavesdropper therefore decodes all
measurements exactly, and all four curves coincide. At $t_c=11$, the user
receives a packet that the eavesdropper misses, and the eavesdropper
loses the decoding reference $\mathbf{y}_{\tau_{t_c}}$. Fig.~\ref{fig:second2} shows that, for arrival probabilities above the critical value $\bar{\delta}_c$, $\mathbb{E}[t_c]$ increases with the transmission energy $E$.
 
\subsection{Attack Detection with the Watermarked CUSUM}\label{subsec:cusum}
 
Fig.~\ref{fig:second} shows one realisation of the CUSUM statistic $S_k$
of Theorem~\ref{thm:cusum_test} when an attack starts at $t_a=400$. Before the
attack, $S_k$ stays near zero, because the log-likelihood ratios
$\ell_t^{g}$ and $\ell_t^{f}$ have negative mean under $\mathcal{H}_0$.
After $t_a$, the mean of the ratio becomes positive. 
 
\subsection{Effect of Transmission Energy on Detectability}\label{subsec:energy}
 
Fig.~\ref{fig:waddE} reports the worst-case average detection delay
 as a function of $E$. WADD is the largest mean detection delay over
any attack start time $t_a$. For the numerical study, we consider all possible values of $t_a$ from 1 to 100. Fig.~\ref{fig:wkldE} shows the
corresponding $\mathrm{WKLD}_f$ (\eqref{eq:wkld}). KLD should be inversely proportional to the WADD. We also observe from the plot that WADD decreases with $E$, whereas WKLD increases. Finally, Fig.~\ref{fig:trace2} shows the trade-off between WADD and
$\mathrm{E}[t_c]$ as the transmission energy $E$ increases.

\subsection{Control Cost of the Watermark}\label{subsec:cost}
 
The watermark covariance $\boldsymbol{\Sigma}_e$ is a second design
parameter. Through $\mathbf{N}=\mathbf{C}\mathbf{B}\boldsymbol{\Sigma}_e
\mathbf{B}^\top\mathbf{C}^\top$, a larger $\boldsymbol{\Sigma}_e$ increases
the trace term in~\eqref{eq:wkld} and hence the WKLD. However, the watermark
acts as an additional process disturbance. Fig.~\ref{fig:energy} shows
the numerically computed closed-loop LQG cost $J$ (evaluated as in~\cite{naha2023quickest}) for different $\boldsymbol{\Sigma}_e$
 and $E$ values. For our numerical study, we only consider diagonal $\boldsymbol{\Sigma}_e$ matrices with equal elements on the main diagonal. Therefore, in the plot, $\boldsymbol{\Sigma}_e$ represents a single element from its main diagonal. The control cost
increases with $\boldsymbol{\Sigma}_e$ but decreases with $E$. Higher  $E$ means more frequent measurement
updates, which in turn improve the state estimate used by the controller and reduce the control cost.



\section{CONCLUSIONS} \label{sec:concl}
We study joint protection of an NCS against eavesdropping and measurement-replacement attacks over lossy wireless channels. The defence mechanism preserves the legitimate receiver's estimation performance during nominal operation, while the eavesdropper's state-estimation error covariance grows without bound after a critical event. We characterise the relationship between the expected time to this critical event and the transmission energy. Higher transmission energy increases the expected time to the critical event, meaning the eavesdropper can reliably reconstruct the system state longer before the critical event occurs. On the other hand, defence against measurement replacement attacks relies on timely detection, characterised by the KLD between the post- and pre-attack distributions of the test signal. However, in the lossy wireless-channel scenario, the conditional KLD depends on the reception history, so it does not admit a simple closed-form expression. Therefore, we derive the worst-case expected KLD as a conservative measure of attack detectability. The WKLD provides a lower bound on expected attack detectability, guiding the design of transmission energy and watermarking strategies to balance confidentiality and timely attack detection.

\section{APPENDIX} \label{sec:appen}
\subsection{Proof of Theorem \ref{thm:cusum_test}}\label{App:the1}
\begingroup
\renewcommand{\theequation}{A.\arabic{equation}}
\setcounter{equation}{0}

\textit{Pre-attack.} Let
$\mathcal{I}_{t}=\{\boldsymbol{\gamma}_{1:t},\mathbf{e}_{1:t}\}$ denote the
information available at time $t$, and recall that the innovation is
$\boldsymbol{\gamma}_t=\mathbf{y}^d_t-\mathbf{C}\hat{\mathbf{x}}_{t|t-1}$.
If $\delta_t^u=1$, then
\begin{equation}
\begin{split}
  \boldsymbol{\gamma}_t
  &=\mathbf{y}_t-\mathbf{C}\hat{\mathbf{x}}_{t|t-1}
   =\mathbf{C}\mathbf{x}_t+\mathbf{v}_t-\mathbf{C}\hat{\mathbf{x}}_{t|t-1}\\
  &=\mathbf{C}\bigl(\mathbf{A}\mathbf{x}_{t-1}
     +\mathbf{B}(\mathbf{L}\hat{\mathbf{x}}_{t-1|t-1}+\mathbf{e}_{t-1})
     +\mathbf{w}_{t-1}\bigr)+\mathbf{v}_t\\
  &\quad-\mathbf{C}\bigl(\mathbf{A}\hat{\mathbf{x}}_{t-1|t-1}
     +\mathbf{B}(\mathbf{L}\hat{\mathbf{x}}_{t-1|t-1}+\mathbf{e}_{t-1})\bigr)\\
  &=\mathbf{C}\mathbf{A}(\mathbf{x}_{t-1}-\hat{\mathbf{x}}_{t-1|t-1})
     +\mathbf{C}\mathbf{w}_{t-1}+\mathbf{v}_t .
\end{split}
\end{equation}
Since the estimation error, $\mathbf{w}_{t-1}$, and $\mathbf{v}_t$ are
zero-mean, $\mathbb{E}[\boldsymbol{\gamma}_t]=\mathbf{0}$. If $\delta^u_t=0$,
then $\boldsymbol{\gamma}_t=\mathbf{0}$. Hence,
$\mathbb{E}[\delta_t^u\boldsymbol{\gamma}_t]=\mathbf{0}$, and, since $\delta_t^u$
is independent of $\boldsymbol{\gamma}_t$,
\begin{equation}
\begin{split}
  \mathrm{cov}(\delta_t^u\boldsymbol{\gamma}_t)
  &=\mathbb{E}[(\delta_t^u)^2\boldsymbol{\gamma}_t\boldsymbol{\gamma}_t^\top]
   =\mathbb{E}[(\delta_t^u)^2]\,
    \mathbb{E}[\boldsymbol{\gamma}_t\boldsymbol{\gamma}_t^\top]\\
  &=\bar{\delta}^u\,\bigl(\mathbf{C}\mathbf{P}_{t|t-1}\mathbf{C}^\top
     +\mathbf{R}\bigr).
\end{split}
\end{equation}

We next show that the innovation sequence is white. By the tower property,
\begin{equation}
  \mathbb{E}[\boldsymbol{\gamma}_t\mid\boldsymbol{\gamma}_{1:t-1},\mathbf{e}_{1:t-1}]
  =\mathbb{E}\bigl[\mathbb{E}[\boldsymbol{\gamma}_t\mid\mathcal{I}_{t-1}]
     \mid\boldsymbol{\gamma}_{1:t-1},\mathbf{e}_{1:t-1}\bigr],
\end{equation}
where
\begin{equation}
\begin{split}
  \mathbb{E}[\boldsymbol{\gamma}_t\mid\mathcal{I}_{t-1}]
  &=\bar{\delta}^u\,\mathbb{E}\bigl[\mathbf{x}_t-\hat{\mathbf{x}}_{t|t-1}
     \mid\mathcal{I}_{t-1}\bigr]\\
  &=\bar{\delta}^u\,\mathbb{E}\bigl[\mathbf{x}_t
     -\mathbb{E}[\mathbf{x}_t\mid\mathcal{I}_{t-1}]\mid\mathcal{I}_{t-1}\bigr]
   =\mathbf{0}.
\end{split}
\end{equation}
Therefore,
$\mathbb{E}[\boldsymbol{\gamma}_t\mid\boldsymbol{\gamma}_{1:t-1},
\mathbf{e}_{1:t-1}]=\mathbf{0}$, i.e., $\boldsymbol{\gamma}_t$ is white.
Finally, by the law of total covariance,
\begin{equation}
\begin{split}
  \mathrm{cov}(\boldsymbol{\gamma}_t\mid\mathcal{I}_{t-1})
  &=\mathbb{E}\bigl[\mathrm{cov}(\boldsymbol{\gamma}_t\mid
     \mathcal{I}_{t-1},\delta^u_t)\bigr]\\
  &\quad+\mathrm{cov}\bigl(\underbrace{\mathbb{E}[\boldsymbol{\gamma}_t
     \mid\mathcal{I}_{t-1},\delta^u_t]}_{=\,\mathbf{0}}\bigr).
\end{split}
\end{equation}
Conditioning on $\delta^u_t=1$ yields
\begin{equation}
  \mathrm{cov}(\boldsymbol{\gamma}_t\mid\mathcal{I}_{t-1},\delta_t^u=1)
  =\mathbf{C}\mathbf{P}_{t|t-1}\mathbf{C}^\top+\mathbf{R}
  =\boldsymbol{\Gamma}_t,
\end{equation}
which establishes~\eqref{eq:cusum_H0}.

\textit{Under attack.} 
Let $t=t_a$ denote the time instant at which the attack starts. Under attack,
the fake measurements evolve according to
\begin{equation}\label{eq:attack_model}
  \mathbf{z}^a_t=\mathbf{A}_a\mathbf{z}^a_{t-1}+\mathbf{w}_{a,t-1},
\end{equation}
and $\bar{\mathbf{z}}_t=\mathbf{z}^a_t$ after the attack. The received signal is
\begin{equation}\label{eq:received}
  \begin{cases}
    \bar{\mathbf{z}}_t, & \delta^u_t=1,\\
    \mathbf{0}, & \delta^u_t=0,
  \end{cases}
\end{equation}
and the decoded measurements are
\begin{equation}\label{eq:decoded}
  \mathbf{y}^d_t=
  \begin{cases}
    \bar{\mathbf{z}}_t+\mathbf{M}^{t-t_k}\mathbf{y}^d_{t_k}, & \delta^u_t=1,\\
    \mathbf{0}, & \delta^u_t=0.
  \end{cases}
\end{equation}
Two cases must be distinguished, depending on the position of $t_k$ relative to
the attack start $t_a$.In both cases, the innovation signal is
\begin{equation}\label{eq:innovation}
  \boldsymbol{\gamma}_t=\mathbf{y}^d_t-\mathbf{C}\hat{\mathbf{x}}_{t|t-1}.
\end{equation}
 
\textbf{Case~1.} We first consider Case~1, i.e., $t_k<t_a$ and $t\ge t_a$. Then
$\mathbf{y}^d_{t_k}=\mathbf{y}_{t_k}$ is the true observation, and
\begin{equation}\label{eq:yd_case1}
  \mathbf{y}^d_t=
  \begin{cases}
    \mathbf{z}^a_t+\mathbf{M}^{t-t_k}\mathbf{y}_t, & \delta_t^u=1,\\
    \mathbf{0}, & \delta_t^u=0.
  \end{cases}
\end{equation}
The innovation becomes
\begin{align}
  \boldsymbol{\gamma}_t
  &=\mathbf{y}^d_t-\mathbf{C}\hat{\mathbf{x}}_{t|t-1}\nonumber\\
  &=\mathbf{z}^a_t+\mathbf{M}^{t-t_k}\mathbf{y}_t
    -\mathbf{C}(\mathbf{A}+\mathbf{B}\mathbf{L})\hat{\mathbf{x}}_{t-1|t-1}
    -\mathbf{C}\mathbf{B}\mathbf{e}_{t-1},\nonumber\\
  &=\mathbf{w}_{a,t-1}+\mathbf{A}_a\mathbf{z}^a_{t-1}
    +\mathbf{M}^{t-t_k}\mathbf{y}_{t_k}\nonumber\\
  &\quad-\mathbf{C}(\mathbf{A}+\mathbf{B}\mathbf{L})\hat{\mathbf{x}}_{t-1|t-1}
    -\mathbf{C}\mathbf{B}\mathbf{e}_{t-1},\label{eq:gamma_case1}
\end{align}
with conditional mean and covariance
\begin{align}
  \mathbb{E}[\boldsymbol{\gamma}_t\mid\mathcal{I}_{\gamma_e}]
  &=\mathbf{M}^{t-t_k}\mathbf{y}_{t_k}
    -\mathbf{C}(\mathbf{A}+\mathbf{B}\mathbf{L})\hat{\mathbf{x}}_{t-1|t-1}
    \nonumber\\
  &\quad-\mathbf{C}\mathbf{B}\mathbf{e}_{t-1},\label{eq:mu_case1}\\
  \mathrm{cov}(\boldsymbol{\gamma}_t\mid\mathcal{I}_{\gamma_e})
  &=\mathbb{E}\bigl[\mathbf{z}^a_t\mathbf{z}^{a\top}_t\bigr]
    =\boldsymbol{\Sigma}_z.\label{eq:Sigma_case1}
\end{align}
Indeed, since
\begin{equation}\label{eq:gamma_tk}
  \boldsymbol{\gamma}_{t_k}=\mathbf{y}_{t_k}-\mathbf{C}\hat{\mathbf{x}}_{t_k|t_k-1},
\end{equation}
$\mathbf{y}_{t_k}$ is known or can be recovered from $\mathcal{I}_{\gamma_e}$.
Moreover, $t_a>t_k$ and $t\ge t_a$ imply that no sample of $\mathbf{z}^a$ has been
received up to time $t-1$. Consequently, $\mathbf{z}^a_{t-1}$ is a random variable
with zero mean and covariance $\boldsymbol{\Sigma}_z$, where $\boldsymbol{\Sigma}_z$
is the solution of the Lyapunov equation
\begin{equation}\label{eq:lyapunov}
  \boldsymbol{\Sigma}_z=\mathbf{A}_a\boldsymbol{\Sigma}_z\mathbf{A}_a^{\top}+\mathbf{Q}_a.
\end{equation}
 
\textbf{Case~2.} We now consider Case~2. For $\delta_t^u=1$,
\begin{align}
  \boldsymbol{\gamma}_t
  &=\mathbf{z}^a_t+\mathbf{M}^{t-t_k}\mathbf{y}^d_{t_k}
    -\mathbf{C}\hat{\mathbf{x}}_{t|t-1}\nonumber\\
  &=\mathbf{w}_{a,t-1}+\mathbf{A}_a\mathbf{z}^a_{t-1}
    +\mathbf{M}^{t-t_k}\mathbf{y}^d_{t_k}\nonumber\\
  &\quad-\mathbf{C}\bigl(\mathbf{A}\hat{\mathbf{x}}_{t-1|t-1}
    +\mathbf{B}\mathbf{L}\hat{\mathbf{x}}_{t-1|t-1}
    +\mathbf{B}\mathbf{e}_{t-1}\bigr)\nonumber\\
  &=\mathbf{w}_{a,t-1}+\mathbf{A}_a\mathbf{z}^a_{t-1}
    +\mathbf{M}^{t-t_k}\mathbf{y}^d_{t_k}\nonumber\\
  &\quad-\mathbf{C}(\mathbf{A}+\mathbf{B}\mathbf{L})\hat{\mathbf{x}}_{t-1|t-1}
    -\mathbf{C}\mathbf{B}\mathbf{e}_{t-1},\label{eq:gamma_case2}
\end{align}
where
\begin{equation}\label{eq:ydtk_case2}
  \mathbf{y}^d_{t_k}=\delta^u_{t_k}\bigl(\boldsymbol{\gamma}_{t_k}
  +\mathbf{C}(\mathbf{A}+\mathbf{B}\mathbf{L})\hat{\mathbf{x}}_{t_k-1|t_k-1}
  +\mathbf{C}\mathbf{B}\mathbf{e}_{t_k-1}\bigr),
\end{equation}
and, for $\delta^u_{t-1}=1$,
\begin{align}
  \boldsymbol{\gamma}_{t-1}
  &=\mathbf{y}^d_{t-1}-\mathbf{C}\hat{\mathbf{x}}_{t-1|t-2}\nonumber\\
  &=\mathbf{z}^a_{t-1}+\mathbf{M}^{t-t_k}\mathbf{y}^d_{t_k}
    -\mathbf{C}\hat{\mathbf{x}}_{t-1|t-2}.\label{eq:gamma_tm1}
\end{align}
If $\delta^u_{t-1}=0$, $\mathbf{z}^a_{t-1}$ cannot be reconstructed. However,
$\delta_{t_k}^u=1$, and iterating~\eqref{eq:attack_model} gives
\begin{equation}\label{eq:za_iter}
  \mathbf{z}^a_t=\mathbf{A}_a^{t-t_k}\mathbf{z}^a_{t_k}
  +\sum_{i=1}^{t-t_k}\mathbf{A}_a^{i-1}\mathbf{w}_{a,t-i}.
\end{equation}
Define the information set
$\mathcal{I}_{\gamma_e}=\{\boldsymbol{\gamma}^{t-1}_1,\mathbf{e}^{t-1}_1\}$.
Substituting~\eqref{eq:za_iter} into~\eqref{eq:gamma_case2} yields
\begin{multline}\label{eq:gamma_case2_expanded}
  \boldsymbol{\gamma}_t=\mathbf{A}_a^{t-t_k}\mathbf{z}^a_{t_k}
  +\sum_{i=1}^{t-t_k}\mathbf{A}_a^{i-1}\mathbf{w}_{a,t-i}
  +\mathbf{M}^{t-t_k}\mathbf{y}^d_{t_k}\\
  -\mathbf{C}(\mathbf{A}+\mathbf{B}\mathbf{L})\hat{\mathbf{x}}_{t-1|t-1}
  -\mathbf{C}\mathbf{B}\mathbf{e}_{t-1}.
\end{multline}
Hence, the conditional mean is
\begin{align}
  \mathbb{E}[\boldsymbol{\gamma}_t\mid\mathcal{I}_{\gamma_e}]
  &=\mathbb{E}\bigl[\boldsymbol{\gamma}_t\mid\hat{\mathbf{x}}_{t-1|t-1},
    \mathbf{e}_{t-1},\mathbf{y}^d_{t_k},\boldsymbol{\gamma}_{t-1}\bigr],
    \nonumber\\
  \boldsymbol{\mu}_{\gamma|\gamma_e}
  &=\mathbf{A}_a^{t-t_k}\mathbf{z}^a_{t_k}
    +\mathbf{M}^{t-t_k}\mathbf{y}^d_{t_k}\nonumber\\
  &\quad-\mathbf{C}(\mathbf{A}+\mathbf{B}\mathbf{L})\hat{\mathbf{x}}_{t-1|t-1}
    -\mathbf{C}\mathbf{B}\mathbf{e}_{t-1}\nonumber\\
  &=\mathbf{A}_a^{t-t_k}\bar{\mathbf{z}}_{t_k}
    +\mathbf{M}^{t-t_k}\mathbf{y}^d_{t_k}\nonumber\\
  &\quad-\mathbf{C}(\mathbf{A}+\mathbf{B}\mathbf{L})\hat{\mathbf{x}}_{t-1|t-1}
    -\mathbf{C}\mathbf{B}\mathbf{e}_{t-1}.\label{eq:mu_case2}
\end{align}
For the conditional covariance,
\begin{multline}\label{eq:cov_def}
  \mathrm{cov}(\boldsymbol{\gamma}_t\mid\mathcal{I}_{\gamma_e})
  =\mathbb{E}\Bigl[\bigl(\boldsymbol{\gamma}_t-\mathbb{E}[\boldsymbol{\gamma}_t\mid\mathcal{I}_{\gamma_e}]\bigr)\\
  \times\bigl(\boldsymbol{\gamma}_t-\mathbb{E}[\boldsymbol{\gamma}_t\mid\mathcal{I}_{\gamma_e}]\bigr)^{\top}
  \Bigm|\mathcal{I}_{\gamma_e}\Bigr],
\end{multline}
where, from~\eqref{eq:gamma_case2_expanded} and~\eqref{eq:mu_case2},
\begin{equation}\label{eq:gamma_dev}
  \boldsymbol{\gamma}_t-\mathbb{E}[\boldsymbol{\gamma}_t\mid\mathcal{I}_{\gamma_e}]
  =\sum_{i=1}^{t-t_k}\mathbf{A}_a^{i-1}\mathbf{w}_{a,t-i}.
\end{equation}
Therefore,
\begin{align}
  \mathrm{cov}(\boldsymbol{\gamma}_t\mid\mathcal{I}_{\gamma_e})
  &=\mathbb{E}\Bigl[\Bigl(\sum_{i=1}^{t-t_k}\mathbf{A}_a^{i-1}\mathbf{w}_{a,t-i}\Bigr)
    \nonumber\\
  &\qquad\times\Bigl(\sum_{i=1}^{t-t_k}\mathbf{A}_a^{i-1}\mathbf{w}_{a,t-i}\Bigr)^{\!\top}\Bigr],
    \nonumber\\
  \boldsymbol{\Sigma}_{\gamma|\gamma_e}
  &=\sum_{i=1}^{t-t_k}\mathbf{A}_a^{i-1}\mathbf{Q}_a\bigl(\mathbf{A}_a^{i-1}\bigr)^{\top}.
  \label{eq:Sigma_case2}
\end{align}
\hfill$\blacksquare$

\endgroup

\subsection{Proof of Theorem \ref{thm:atbce}}\label{App:the2}

\begingroup
\renewcommand{\theequation}{B.\arabic{equation}}
\setcounter{equation}{0}

By Assumption~\ref{ass:reception}, at each time instant $t=1,2,\dots$, the
critical event $\mathcal{C}_t$ occurs with probability
\begin{equation}
  \mathbb{P}\{\mathcal{C}_t\}
  = \mathbb{P}\{\delta^u_t = 1\}\,\mathbb{P}\{\delta^e_t = 0\}
  = \bar\delta_u\,(1-\bar\delta_e) = p_c,
  \label{eq:B_pc}
\end{equation}
independently across time.
Since $t_c=t$ if and only if $\mathcal{C}_1,\dots,\mathcal{C}_{t-1}$ do not
occur and $\mathcal{C}_t$ occurs,
\begin{equation}
  \mathbb{P}\{t_c = t\} = (1 - p_c)^{t-1}\,p_c,
  \qquad t = 1, 2, \dots,
  \label{eq:B_geom}
\end{equation}
that is, $t_c$ is geometrically distributed with success probability
$p_c\in(0,1)$; in particular, $t_c<\infty$ almost surely. Consequently,
\begin{equation}
  \begin{split}
    \mathbb{E}[t_c]
    &= \sum_{t=1}^{\infty} t\,(1 - p_c)^{t-1}\,p_c
     = \frac{p_c}{\bigl(1-(1-p_c)\bigr)^2} \\
    &= \frac{1}{p_c}
     = \frac{1}{\bar\delta_u\,(1 - \bar\delta_e)},
  \end{split}
  \label{eq:B_mean}
\end{equation}
where we used $\sum_{t\ge1} t\,x^{t-1} = (1-x)^{-2}$ for $|x|<1$.
This proves \eqref{eq:atbce}.

Finally, let $\bar\delta_u=\bar\delta_e=:\bar\delta$. By \eqref{eq:energy},
$\bar\delta=1-\epsilon^E$, and substituting into \eqref{eq:B_mean} gives
\begin{equation}
  \mathbb{E}[t_c]=\frac{1}{\bar\delta(1-\bar\delta)}
  =\frac{1}{\epsilon^E(1-\epsilon^E)}.
  \label{eq:B_eps}
\end{equation}

\endgroup

\subsection{Proof of Lemma \ref{lem:eaves}(ii)}\label{App:lem_eaves}

\edef\savedeqnumD{\arabic{equation}}
\setcounter{equation}{0}
\renewcommand{\theequation}{D.\arabic{equation}}

Let $\boldsymbol{\Pi}_{t|t-1}$ be the covariance computed by the eavesdropper's
Kalman filter, i.e., the Riccati recursion~\eqref{eq:kfP}--\eqref{eq:kfupd} with
$\delta_t$ replaced by $\delta^e_t$, let
$\mathbf{K}^e_t=\boldsymbol{\Pi}_{t|t-1}\mathbf{C}^\top(\mathbf{C}\boldsymbol{\Pi}_{t|t-1}\mathbf{C}^\top+\mathbf{R})^{-1}$,
$\mathbf{F}_t:=\mathbf{I}-\mathbf{K}^e_t\mathbf{C}$, and
$\tilde{\mathbf{x}}^e_{t|s}:=\mathbf{x}_t-\hat{\mathbf{x}}^e_{t|s}$.

\emph{Step 1 (decoding mismatch).} Let
$\Delta\mathbf{y}_t:=\mathbf{y}_t-\mathbf{y}^e_t$ and
\begin{equation}\label{eq:D_eta}
  \boldsymbol{\eta}:=\mathbf{y}_{t_c}-\mathbf{C}\hat{\mathbf{x}}^e_{t_c|t_c-1}
  =\mathbf{C}\tilde{\mathbf{x}}^e_{t_c|t_c-1}+\mathbf{v}_{t_c}.
\end{equation}
For every user-reception time $s\ge t_c$, let
$\boldsymbol{\varepsilon}_s:=\mathbf{y}_s-\mathbf{r}^e_s$. By
Assumption~\ref{ass:eaves}, $\mathbf{r}^e_{t_c}=\mathbf{C}\hat{\mathbf{x}}^e_{t_c|t_c-1}$,
so $\boldsymbol{\varepsilon}_{t_c}=\boldsymbol{\eta}$. For $s>t_c$ with
$\delta^u_s=1$, we have $\delta^e_s=1$, so
$\mathbf{r}^e_s=\mathbf{y}^e_s=\mathbf{z}_s+\mathbf{M}^{k_s}\mathbf{r}^e_{\tau_s}$,
whereas~\eqref{eq:encoder} gives
$\mathbf{y}_s=\mathbf{z}_s+\mathbf{M}^{k_s}\mathbf{y}_{\tau_s}$. Hence
$\boldsymbol{\varepsilon}_s=\mathbf{M}^{k_s}\boldsymbol{\varepsilon}_{\tau_s}$, and
since $k_s=s-\tau_s$, induction gives
$\boldsymbol{\varepsilon}_s=\mathbf{M}^{s-t_c}\boldsymbol{\eta}$. For any $t>t_c$,
$\tau_t\ge t_c$, and therefore
\begin{equation}\label{eq:D_dy}
  \Delta\mathbf{y}_t=\mathbf{M}^{k_t}\boldsymbol{\varepsilon}_{\tau_t}
  =\mathbf{M}^{t-t_c}\boldsymbol{\eta}.
\end{equation}

\emph{Step 2 (error dynamics).} The eavesdropper knows the applied input
$\mathbf{u}_{t-1}+\mathbf{e}_{t-1}$, so
$\tilde{\mathbf{x}}^e_{t|t-1}=\mathbf{A}\tilde{\mathbf{x}}^e_{t-1|t-1}+\mathbf{w}_{t-1}$.
For $t>t_c$, $\delta^e_t=1$ and
$\mathbf{y}^e_t=\mathbf{C}\mathbf{x}_t+\mathbf{v}_t-\Delta\mathbf{y}_t$, so
by~\eqref{eq:D_dy}
\begin{equation}\label{eq:D_upd}
  \tilde{\mathbf{x}}^e_{t|t}
  =\mathbf{F}_t\mathbf{A}\tilde{\mathbf{x}}^e_{t-1|t-1}+\mathbf{F}_t\mathbf{w}_{t-1}
   -\mathbf{K}^e_t\mathbf{v}_t+\mathbf{K}^e_t\mathbf{M}^{t-t_c}\boldsymbol{\eta}.
\end{equation}

\emph{Step 3 (lower bound).} For $t<t_c$, every user reception is also an
eavesdropper reception, so the eavesdropper decodes exactly and its filter is matched.
Hence $\tilde{\mathbf{x}}^e_{t_c|t_c-1}$ has covariance
$\mathbf{P}_c:=\boldsymbol{\Pi}_{t_c|t_c-1}$, and, since $\delta^e_{t_c}=0$,
$\tilde{\mathbf{x}}^e_{t_c|t_c}=\tilde{\mathbf{x}}^e_{t_c|t_c-1}$. As $\mathbf{v}_{t_c}$
is independent of $\tilde{\mathbf{x}}^e_{t_c|t_c-1}$,~\eqref{eq:D_eta} gives
$\mathbf{H}:=\mathbb{E}[\boldsymbol{\eta}\boldsymbol{\eta}^\top]=\mathbf{C}\mathbf{P}_c\mathbf{C}^\top+\mathbf{R}\succeq\mathbf{R}\succ0$.
Define
$\boldsymbol{\Psi}_t:=\mathbb{E}[\tilde{\mathbf{x}}^e_{t|t}\boldsymbol{\eta}^\top]\mathbf{H}^{-1}$.
Since the second-moment matrix of
$(\tilde{\mathbf{x}}^e_{t|t},\boldsymbol{\eta})$ is positive semidefinite, its Schur
complement gives
$\mathbf{P}^e_{t|t}\succeq\boldsymbol{\Psi}_t\mathbf{H}\boldsymbol{\Psi}_t^\top\succeq\lambda_{\min}(\mathbf{R})\boldsymbol{\Psi}_t\boldsymbol{\Psi}_t^\top$.
Let $\lambda\in\mathbb{C}$ be an eigenvalue of $\mathbf{M}$ with
$|\lambda|=\rho(\mathbf{M})>1$, and let $\mathbf{v}$ be a corresponding eigenvector with
$\|\mathbf{v}\|=1$. Then
\begin{equation}\label{eq:D_black}
  \operatorname{tr}\bigl(\mathbf{P}^e_{t|t}\bigr)
  \ge\lambda_{\min}(\mathbf{R})\,\|\boldsymbol{\Psi}_t\|_F^2
  \ge\lambda_{\min}(\mathbf{R})\,\|\boldsymbol{\Psi}_t\mathbf{v}\|^2 .
\end{equation}

\emph{Step 4 (recursion for $\boldsymbol{\Psi}_t$).} For $t>t_c$, the noises
$\mathbf{w}_{t-1}$ and $\mathbf{v}_t$ are independent of $\boldsymbol{\eta}$, and the
gains are deterministic for a given reception path. Multiplying~\eqref{eq:D_upd} by
$\boldsymbol{\eta}^\top$, taking expectations and multiplying by $\mathbf{H}^{-1}$
gives
\begin{equation*}
  \boldsymbol{\Psi}_t=\mathbf{F}_t\mathbf{A}\boldsymbol{\Psi}_{t-1}+\mathbf{K}^e_t\mathbf{M}^{t-t_c},
  \qquad \boldsymbol{\Psi}_{t_c}=\mathbf{P}_c\mathbf{C}^\top\mathbf{H}^{-1}.
\end{equation*}
Let $\mathbf{r}_t:=\lambda^{-(t-t_c)}\boldsymbol{\Psi}_t\mathbf{v}$. Since
$\mathbf{M}^{t-t_c}\mathbf{v}=\lambda^{t-t_c}\mathbf{v}$,
\begin{equation}\label{eq:D_r}
  \mathbf{r}_t=\lambda^{-1}\mathbf{F}_t\mathbf{A}\,\mathbf{r}_{t-1}+\mathbf{K}^e_t\mathbf{v}.
\end{equation}

\emph{Step 5 (limit).} For $t>t_c$, $\delta^e_t=1$, so $\boldsymbol{\Pi}_{t|t-1}$
follows the standard Riccati recursion. By detectability and stabilisability,
$\boldsymbol{\Pi}_{t|t-1}\to\bar{\boldsymbol{\Pi}}$, the stabilising solution of the
algebraic Riccati equation~\cite{anderson1979optimal}. Hence
$\mathbf{K}^e_t\to\bar{\mathbf{K}}:=\bar{\boldsymbol{\Pi}}\mathbf{C}^\top(\mathbf{C}\bar{\boldsymbol{\Pi}}\mathbf{C}^\top+\mathbf{R})^{-1}$
and $\mathbf{F}_t\mathbf{A}\to\bar{\mathbf{F}}\mathbf{A}$, where
$\bar{\mathbf{F}}:=\mathbf{I}-\bar{\mathbf{K}}\mathbf{C}$ and
$\rho(\bar{\mathbf{F}}\mathbf{A})<1$. Since $|\lambda|>1$,
$\lambda^{-1}\bar{\mathbf{F}}\mathbf{A}$ is Schur stable, and~\eqref{eq:D_r} is an
asymptotically time-invariant, exponentially stable recursion with a convergent input.
Therefore $\mathbf{r}_t\to\bar{\mathbf{r}}:=(\mathbf{I}-\lambda^{-1}\bar{\mathbf{F}}\mathbf{A})^{-1}\bar{\mathbf{K}}\mathbf{v}$.

\emph{Step 6 ($\bar{\mathbf{r}}\neq\mathbf{0}$).} Since
$\bar{\boldsymbol{\Pi}}\succeq\mathbf{Q}$ and $\mathbf{C}\mathbf{Q}\mathbf{C}^\top\succ0$,
the matrix
$\mathbf{C}\bar{\mathbf{K}}=\mathbf{C}\bar{\boldsymbol{\Pi}}\mathbf{C}^\top(\mathbf{C}\bar{\boldsymbol{\Pi}}\mathbf{C}^\top+\mathbf{R})^{-1}$
is a product of two positive definite matrices and hence invertible. Thus
$\bar{\mathbf{K}}\mathbf{v}\neq\mathbf{0}$ and $\bar{\mathbf{r}}\neq\mathbf{0}$.

\emph{Step 7 (conclusion).} By Step~5, there is a finite $T_0$ such that
$\|\mathbf{r}_t\|\ge\tfrac12\|\bar{\mathbf{r}}\|$ for $t\ge t_c+T_0$. Since
$\|\boldsymbol{\Psi}_t\mathbf{v}\|=\rho(\mathbf{M})^{t-t_c}\|\mathbf{r}_t\|$,
\eqref{eq:D_black} gives
\begin{equation*}
  \operatorname{tr}\bigl(\mathbf{P}^e_{t|t}\bigr)\ge c\,\rho(\mathbf{M})^{2(t-t_c)},\qquad
  c:=\tfrac14\lambda_{\min}(\mathbf{R})\,\|\bar{\mathbf{r}}\|^2,
\end{equation*}
for all $t\ge t_c+T_0$. The constant $c$ depends only on
$(\mathbf{A},\mathbf{C},\mathbf{Q},\mathbf{R},\mathbf{M})$, not on $t_c$. Since
$\rho(\mathbf{M})>1$, $\operatorname{tr}(\mathbf{P}^e_{t|t})\to\infty$.
\hfill$\blacksquare$

\setcounter{equation}{\savedeqnumD}

\subsection{Proof of Theorem \ref{thm:wkld}}\label{App:the3}

\edef\savedeqnum{\arabic{equation}}
\begingroup
\setcounter{equation}{0}
\renewcommand{\theequation}{C.\arabic{equation}}
The KLD between two Gaussian distributions $\mathcal{N}(\boldsymbol{\mu}_1,\boldsymbol{\Sigma}_1)$
and $\mathcal{N}(\boldsymbol{\mu}_0,\boldsymbol{\Sigma}_0)$ is given by \cite{naha2023quickest}
\begin{multline}
D_g = \frac{1}{2}\Big[ \operatorname{tr}\big(\boldsymbol{\Sigma}_0^{-1}\boldsymbol{\Sigma}_1\big) - p
 + (\boldsymbol{\mu}_1-\boldsymbol{\mu}_0)^\top \boldsymbol{\Sigma}_0^{-1}(\boldsymbol{\mu}_1-\boldsymbol{\mu}_0) \\
 + \log\frac{\det \boldsymbol{\Sigma}_0}{\det \boldsymbol{\Sigma}_1} \Big]. \tag{C.1}
\end{multline}
We consider the case $j=f$ ($t_a \le \tau_t$)\textcolor{black}{. The densities compared by the
CUSUM test~\eqref{eq:cusum_llr} are the conditional densities of the innovation
given $\mathcal{I}_{t-1}$ \emph{and} $\delta^u_t=1$, see~\eqref{eq:cusum_H1}
and~\eqref{eq:cusum_H0}. Hence, on $\{\delta^u_t=1\}$,} $\boldsymbol{\mu}_0 = \mathbf{0}$,
\textcolor{black}{$\boldsymbol{\Sigma}_0 = \boldsymbol{\Gamma}_t
 = \mathbf{C}\mathbf{P}_{t|t-1}\mathbf{C}^\top + \mathbf{R}$ from~\eqref{eq:cusum_H0}},
$\boldsymbol{\mu}_1 = \boldsymbol{\mu}^f_t$ from~\eqref{eq:cusum_mu}, and
$\boldsymbol{\Sigma}_1 = \boldsymbol{\Lambda}_{k_t}$ from~\eqref{eq:cusum_Sig}.
\textcolor{black}{(The covariance $\bar{\delta}^u\boldsymbol{\Gamma}_t$ in (A.2) is the
unconditional covariance of $\delta^u_t\boldsymbol{\gamma}_t$, which is a Gaussian
mixture with a point mass at $\mathbf{0}$ and is not the density used in the test.)}
These parameters depend on the
past information $I_{t-1} = \{\bar{\gamma}_{1:t-1}, \mathbf{e}_{1:t-1}, \delta_{1:t-1}\}$:
$\boldsymbol{\mu}_1$ through the past innovations, the watermark and the last received
falsified packet $\bar{\mathbf{z}}_{\tau_t}$, while $\mathbf{P}_{t|t-1}$ and $k_t$, and hence
$\boldsymbol{\Lambda}_{k_t}$, depend on the packet-reception history. Hence the conditional KLD
$D\big(f^f_{\tilde{\gamma}_t}, f_{\gamma_t} \mid I_{t-1}\big)$, given by $D_g$ in (C.1), is a
random variable, and, as in \cite{naha2023quickest}, we take its expectation over the joint distribution
of $I_{t-1}$. Since $\boldsymbol{\mu}_1^\top\boldsymbol{\Sigma}_0^{-1}\boldsymbol{\mu}_1$ is a scalar,
the cyclic property of the trace gives
$\boldsymbol{\mu}_1^\top\boldsymbol{\Sigma}_0^{-1}\boldsymbol{\mu}_1
 = \operatorname{tr}(\boldsymbol{\Sigma}_0^{-1}\boldsymbol{\mu}_1\boldsymbol{\mu}_1^\top)$.
\textcolor{black}{When $\delta^u_t=0$, no packet is received, the statistic~\eqref{eq:cusum_W}
is not updated, and the conditional KLD is zero by convention. Since $\delta^u_t$ is
independent of $I_{t-1}$ and $\mathbb{P}\{\delta^u_t=1\}=\bar{\delta}^u$,}
by linearity of the trace, the trace and mean terms combine as
\begin{multline}
\mathbb{E}\big[D\big(f^f_{\tilde{\gamma}_t}, f_{\gamma_t} \mid I_{t-1}\big)\big]
 = \frac{\bar{\delta}^u}{2}\,\mathbb{E}\Big[
 \operatorname{tr}\big(\boldsymbol{\Gamma}_t^{-1}(\boldsymbol{\Lambda}_{k_t}
 + \boldsymbol{\mu}_1\boldsymbol{\mu}_1^\top)\big) \\
 - p + \log\frac{\det\boldsymbol{\Gamma}_t}{\det\boldsymbol{\Lambda}_{k_t}} \Big]. \tag{C.2}
\end{multline}

\emph{Mean term.} By~\eqref{eq:cusum_mu}, we decompose
$\boldsymbol{\mu}_1 = \boldsymbol{\xi}_t + \mathbf{A}_a^{k_t}\bar{\mathbf{z}}_{\tau_t} - \mathbf{C}\mathbf{B}\mathbf{e}_{t-1}$,
where $\boldsymbol{\xi}_t = \mathbf{M}^{k_t}\mathbf{y}^d_{\tau_t}
 - \mathbf{C}(\mathbf{A}+\mathbf{B}\mathbf{L})\hat{\mathbf{x}}_{t-1|t-1}$.
The terms of $\boldsymbol{\xi}_t$ depend on the watermark only up to $\mathbf{e}_{t-2}$,
the attack process is independent of the watermark, and $\boldsymbol{\Gamma}_t$ and $k_t$
depend only on the packet-reception history. Since $\mathbf{e}_{t-1}$ is zero mean and
independent of $\boldsymbol{\xi}_t$, $\mathbf{A}_a^{k_t}\bar{\mathbf{z}}_{\tau_t}$ and
$\boldsymbol{\Gamma}_t$, the cross terms vanish, and, with
$\mathbf{N} := \mathbf{C}\mathbf{B}\boldsymbol{\Sigma}_e\mathbf{B}^\top\mathbf{C}^\top \succeq 0$,
\begin{multline}
\mathbb{E}\big[\operatorname{tr}\big(\boldsymbol{\Gamma}_t^{-1}\boldsymbol{\mu}_1\boldsymbol{\mu}_1^\top\big)\big]\\
 = \mathbb{E}\big[\operatorname{tr}\big(\boldsymbol{\Gamma}_t^{-1}
 (\boldsymbol{\xi}_t + \mathbf{A}_a^{k_t}\bar{\mathbf{z}}_{\tau_t})
 (\boldsymbol{\xi}_t + \mathbf{A}_a^{k_t}\bar{\mathbf{z}}_{\tau_t})^\top\big)\big] \\
 + \mathbb{E}\big[\operatorname{tr}\big(\boldsymbol{\Gamma}_t^{-1}\mathbf{N}\big)\big]
 \ge \mathbb{E}\big[\operatorname{tr}\big(\boldsymbol{\Gamma}_t^{-1}\mathbf{N}\big)\big]. \tag{C.3}
\end{multline}
Hence,  the watermark guarantees a non-zero mean shift under
attack, even when the decoding reference $\bar{\mathbf{z}}_{\tau_t}$ is falsified.

\emph{Bounds on $\boldsymbol{\Lambda}_{k_t}$.} Since $\delta^u_t = 1$, we have $k_t \ge 1$.
From~\eqref{eq:cusum_Sig}, $\boldsymbol{\Lambda}_{k} = \mathbf{Q}_a
 + \sum_{i=1}^{k-1}\mathbf{A}_a^i\mathbf{Q}_a(\mathbf{A}_a^i)^\top \succeq \mathbf{Q}_a$, and,
since $\rho(\mathbf{A}_a)<1$, $\boldsymbol{\Sigma}_z = \sum_{i=0}^{\infty}
 \mathbf{A}_a^i\mathbf{Q}_a(\mathbf{A}_a^i)^\top$, so that
$\boldsymbol{\Sigma}_z - \boldsymbol{\Lambda}_k
 = \mathbf{A}_a^{k}\boldsymbol{\Sigma}_z(\mathbf{A}_a^{k})^\top \succeq 0$. Therefore,
for every realisation of $k_t$,
\begin{equation}
\mathbf{Q}_a \preceq \boldsymbol{\Lambda}_{k_t} \preceq \boldsymbol{\Sigma}_z. \tag{C.4}
\end{equation}

\emph{Trace term.} Substituting (C.3) into (C.2), and using the left inequality in (C.4)
together with $\operatorname{tr}(\mathbf{X}^{-1}\mathbf{Y}) \ge 0$ for $\mathbf{X}\succ 0$,
$\mathbf{Y}\succeq 0$, gives
$\mathbb{E}[\operatorname{tr}(\boldsymbol{\Gamma}_t^{-1}(\boldsymbol{\Lambda}_{k_t} + \mathbf{N}))]
 \ge \mathbb{E}[\operatorname{tr}(\boldsymbol{\Gamma}_t^{-1}(\mathbf{Q}_a + \mathbf{N}))]$,
in which the second argument is now deterministic. Since
$\mathbf{X} \mapsto \operatorname{tr}(\mathbf{X}^{-1}\mathbf{Y})$ is convex
\textcolor{black}{and nonincreasing (in the Loewner order) on $\mathbf{X}\succ0$} for
$\mathbf{Y}\succeq 0$, Jensen's inequality and, in steady state,
\textcolor{black}{$\mathbb{E}[\boldsymbol{\Gamma}_t]
 = \mathbf{C}\,\mathbb{E}[\mathbf{P}_{t|t-1}]\mathbf{C}^\top+\mathbf{R}
 \preceq \mathbf{C}\bar{\mathbf{P}}\mathbf{C}^\top+\mathbf{R}$~\cite{schenato2007foundations}} yield
\begin{multline}
\mathbb{E}\big[\operatorname{tr}\big(\boldsymbol{\Gamma}_t^{-1}(\boldsymbol{\Lambda}_{k_t}+\mathbf{N})\big)\big]
 \ge \operatorname{tr}\big(\mathbb{E}[\boldsymbol{\Gamma}_t]^{-1}(\mathbf{Q}_a+\mathbf{N})\big)\\
 \ge \operatorname{tr}\big((\mathbf{C}\bar{\mathbf{P}}\mathbf{C}^\top + \mathbf{R})^{-1}
 (\mathbf{Q}_a+\mathbf{N})\big). \tag{C.5}
\end{multline}

\emph{Log-determinant term.} By the right inequality in (C.4),
$\log\det\boldsymbol{\Lambda}_{k_t} \le \log\det\boldsymbol{\Sigma}_z$
\textcolor{black}{for every realisation of $k_t$. For $\log\det\boldsymbol{\Gamma}_t$ a
\emph{lower} bound is required. Since $\log\det$ is concave, Jensen's inequality
only gives the upper bound
$\mathbb{E}[\log\det\boldsymbol{\Gamma}_t]\le\log\det\mathbb{E}[\boldsymbol{\Gamma}_t]$,
so a bound on $\mathbb{E}[\mathbf{P}_{t|t-1}]$ cannot be used here. Instead, we use an
almost-sure lower bound on $\mathbf{P}_{t|t-1}$. Let
$g_{\delta}(\mathbf{P}) := \mathbf{A}\mathbf{P}\mathbf{A}^\top+\mathbf{Q}
 - \delta\,\mathbf{A}\mathbf{P}\mathbf{C}^\top(\mathbf{C}\mathbf{P}\mathbf{C}^\top+\mathbf{R})^{-1}\mathbf{C}\mathbf{P}\mathbf{A}^\top$,
so that, by~\eqref{eq:kfP}--\eqref{eq:kfupd},
$\mathbf{P}_{t|t-1}=g_{\delta^u_{t-1}}(\mathbf{P}_{t-1|t-2})$. For every
$\mathbf{P}\succeq0$, $g_1(\mathbf{P})\preceq g_0(\mathbf{P})$, and $g_1$ is monotone
in the Loewner order~\cite{schenato2007foundations}. Let
$\mathbf{P}^{\star}_{t|t-1}:=g_1(\mathbf{P}^{\star}_{t-1|t-2})$ with
$\mathbf{P}^{\star}_{1|0}=\mathbf{P}_{1|0}$ be the lossless Riccati recursion. By
induction, $\mathbf{P}_{t|t-1}=g_{\delta^u_{t-1}}(\mathbf{P}_{t-1|t-2})
 \succeq g_1(\mathbf{P}_{t-1|t-2})\succeq g_1(\mathbf{P}^{\star}_{t-1|t-2})
 =\mathbf{P}^{\star}_{t|t-1}$ for every packet-reception realisation. Under the
detectability and stabilisability assumptions,
$\mathbf{P}^{\star}_{t|t-1}\to\mathbf{P}^{\star}$, the stabilising solution of the
standard ARE~\cite{anderson1979optimal}. Hence, in steady state,
$\boldsymbol{\Gamma}_t\succeq\mathbf{C}\mathbf{P}^{\star}\mathbf{C}^\top+\mathbf{R}$
almost surely, and by monotonicity of $\log\det$,}
\begin{equation}
\mathbb{E}\Big[\log\frac{\det\boldsymbol{\Gamma}_t}{\det\boldsymbol{\Lambda}_{k_t}}\Big]
 \ge \log\frac{\big|\mathbf{C}\mathbf{P}^{\star}\mathbf{C}^\top+\mathbf{R}\big|}
 {|\boldsymbol{\Sigma}_z|}. \tag{C.6}
\end{equation}
Applying (C.5) and (C.6) in (C.2) yields
\begin{multline*}
\mathbb{E}\big[D\big(f^f_{\tilde{\gamma}_t}, f_{\gamma_t} \mid I_{t-1}\big)\big]\\
 \ge \frac{\bar{\delta}^u}{2}\Big(
 \operatorname{tr}\big((\mathbf{C}\bar{\mathbf{P}}\mathbf{C}^\top+\mathbf{R})^{-1}
 (\mathbf{Q}_a+\mathbf{N})\big) \\
 - p + \log\frac{\big|\mathbf{C}\mathbf{P}^{\star}\mathbf{C}^\top+\mathbf{R}\big|}
 {|\boldsymbol{\Sigma}_z|}\Big),
\end{multline*}
which is~\eqref{eq:wkld}. \hfill$\blacksquare$

\endgroup
\setcounter{equation}{\savedeqnum}

\subsection{Simulation Parameters}\label{apdx:num_parameters}
The plant is a discretised ($T_s = 0.1$\,s) three-node network with
\begin{equation*}
A = \begin{bmatrix} I_3 & T_s I_3 \\
 -T_s M_g^{-1}\!\left(L_{gl}L_{ll}^{-1}L_{lg} - L_{gg}\right) & I_3 - T_s M_g^{-1} D_g \end{bmatrix},\end{equation*}
 \begin{equation*}
B = \begin{bmatrix} 0_3 \\ T_s M_g^{-1} \end{bmatrix},\quad
C = \begin{bmatrix} I_3 & I_3 \end{bmatrix},
\end{equation*}
where $M_g = \mathrm{diag}(0.125, 0.034, 0.016)$,
$D_g = \mathrm{diag}(0.125, 0.068, 0.48)$,
$L_{gg} = \mathrm{diag}(0.058, 0.063, 0.059)$,
$L_{gl} = [-L_{gg}\;\; 0_3]$, $L_{lg} = L_{gl}^\top$, and
\begin{equation*}
L_{ll} = {\setlength{\arraycolsep}{2pt}\left[\begin{array}{@{}cccccc@{}}
 \scriptstyle 0.235 & \scriptstyle 0 & \scriptstyle 0 & \scriptstyle -0.085 & \scriptstyle -0.092 & \scriptstyle 0\\
 \scriptstyle 0 & \scriptstyle 0.296 & \scriptstyle 0 & \scriptstyle -0.161 & \scriptstyle 0 & \scriptstyle -0.072\\
 \scriptstyle 0 & \scriptstyle 0 & \scriptstyle 0.33 & \scriptstyle 0 & \scriptstyle -0.17 & \scriptstyle -0.101\\
 \scriptstyle -0.085 & \scriptstyle -0.161 & \scriptstyle 0 & \scriptstyle 0.246 & \scriptstyle 0 & \scriptstyle 0\\
 \scriptstyle -0.092 & \scriptstyle 0 & \scriptstyle -0.17 & \scriptstyle 0 & \scriptstyle 0.262 & \scriptstyle 0\\
 \scriptstyle 0 & \scriptstyle -0.072 & \scriptstyle -0.101 & \scriptstyle 0 & \scriptstyle 0 & \scriptstyle 0.173
\end{array}\right]}.
\end{equation*}
The system is open-loop unstable with $\rho(A) \approx 2.006$.
The noise covariances are $Q = I_6$ and $R = I_3$, and the initial state is drawn from
$x_0 \sim \mathcal{N}(0, 0.1 I_6)$.

\bibliographystyle{IEEEtran}
\bibliography{ECC}

\end{document}